\documentclass[final]{agujournal2019}
\usepackage{url} %this package should fix any errors with URLs in refs.
\usepackage{lineno}
\usepackage[inline]{trackchanges} %for better track changes. finalnew option will compile document with changes incorporated.
\usepackage{soul}
\usepackage{amssymb}
\usepackage[final]{listings} % use "final" to get over blank outputs in "draft" mode
\usepackage{textcomp}
\definecolor{grey}{rgb}{0.5,0.5,0.5}
\definecolor{mauve}{rgb}{0.58,0,0.82}
\definecolor{darkgreen}{rgb}{0,0.3,0}
\definecolor{darkred}{rgb}{0.8,0,0}
\draftfalse

\journalname{Journal of Advances in Modeling Earth Systems (JAMES)}

\begin{document}

%% ------------------------------------------------------------------------ %%
%  Title
%
% (A title should be specific, informative, and brief. Use
% abbreviations only if they are defined in the abstract. Titles that
% start with general keywords then specific terms are optimized in
% searches)
%
%% ------------------------------------------------------------------------ %%

% Example: \title{This is a test title}

\title{On constraining the mesoscale eddy energy dissipation time-scale}

%% ------------------------------------------------------------------------ %%
%
%  AUTHORS AND AFFILIATIONS
%
%% ------------------------------------------------------------------------ %%

% Authors are individuals who have significantly contributed to the
% research and preparation of the article. Group authors are allowed, if
% each author in the group is separately identified in an appendix.)

% List authors by first name or initial followed by last name and
% separated by commas. Use \affil{} to number affiliations, and
% \thanks{} for author notes.
% Additional author notes should be indicated with \thanks{} (for
% example, for current addresses).

% Example: \authors{A. B. Author\affil{1}\thanks{Current address, Antartica}, B. C. Author\affil{2,3}, and D. E.
% Author\affil{3,4}\thanks{Also funded by Monsanto.}}

\authors{J. Mak\affil{1,2}, A. Avdis\affil{3,4}, T. W. David\affil{5}, H. S. Lee\affil{1}, Y. Na\affil{1}, Y. Wang\affil{1,2} and F. E. Yan\affil{1}}

% \affiliation{1}{First Affiliation}
% \affiliation{2}{Second Affiliation}
% \affiliation{3}{Third Affiliation}
% \affiliation{4}{Fourth Affiliation}

\affiliation{1}{Department of Ocean Science, Hong Kong University of Science and Technology}
\affiliation{2}{Center for Ocean Research in Hong Kong and Macau, Hong Kong University of Science and Technology}
\affiliation{3}{Department of Earth Science and Engineering, Imperial College London}
\affiliation{4}{HPC department, Boston Limited}
\affiliation{5}{Laboratoire de Physique, \'Ecole Normale Sup\'erieure de Lyon}
%(repeat as many times as is necessary)

%% Corresponding Author:
% Corresponding author mailing address and e-mail address:

% (include name and email addresses of the corresponding author.  More
% than one corresponding author is allowed in this LaTeX file and for
% publication; but only one corresponding author is allowed in our
% editorial system.)

% Example: \correspondingauthor{First and Last Name}{email@address.edu}

\correspondingauthor{Julian Mak}{julian.c.l.mak@googlemail.com}

%% Keypoints, final entry on title page.

%  List up to three key points (at least one is required)
%  Key Points summarize the main points and conclusions of the article
%  Each must be 140 characters or fewer with no special characters or punctuation and must be complete sentences

% Example:
% \begin{keypoints}
% \item	List up to three key points (at least one is required)
% \item	Key Points summarize the main points and conclusions of the article
% \item	Each must be 140 characters or fewer with no special characters or punctuation and must be complete sentences
% \end{keypoints}

\begin{keypoints}
\item Constraining the mesoscale eddy energy dissipation time-scale in the ocean
via a simple and computationally inexpensive inverse calculation
\item Shortest dissipation time-scale in Southern Ocean, Western Boundary
Currents, and ocean western boundaries
\item Computes a spatially varying lower bound dissipation time-scale to guide
future investigations

%\begin{verbatim}
%          ================================================
%                |\      _,,,---,,_
%          ZZZzz /,`.-'`'    -.  ;-;;,_
%               |,4-  ) )-,_. ,\ (  `'-'
%              '---''(_/--'  `-'\_)         Here, Miffy was
%          ================================================
%\end{verbatim}
\end{keypoints}

%% ------------------------------------------------------------------------ %%
%
%  ABSTRACT and PLAIN LANGUAGE SUMMARY
%
% A good Abstract will begin with a short description of the problem
% being addressed, briefly describe the new data or analyses, then
% briefly states the main conclusion(s) and how they are supported and
% uncertainties.

% The Plain Language Summary should be written for a broad audience,
% including journalists and the science-interested public, that will not have 
% a background in your field.
%
% A Plain Language Summary is required in GRL, JGR: Planets, JGR: Biogeosciences,
% JGR: Oceans, G-Cubed, Reviews of Geophysics, and JAMES.
% see http://sharingscience.agu.org/creating-plain-language-summary/)
%
%% ------------------------------------------------------------------------ %%

%% \begin{abstract} starts the second page

\begin{abstract}
A physically plausible lower bound on the spatially varying geostrophic
mesoscale eddy energy dissipation time-scale within the ocean, related to the
geographical energy transfer rate out of the geostrophic mesoscales, is provided
by means of a simple and computational inexpensive inverse calculation. Data
diagnosed from a high resolution global configuration ocean simulation is
supplied to a parameterized model of the geostrophic mesoscale eddy energy, from
which the dissipation time-scale results as a solution to an optimization
calculation. We find that the dissipation time-scale is shortest in the Southern
Ocean, in the Western Boundary Currents, and on the western boundaries,
consistent with the expectation that these regions are notable sites of
baroclinic activity with processes leading to energy transfer out of the
geostrophic mesoscales. Although our solution should be interpreted as a lower
bound given the assumptions going into the calculation, it serves as an
important physically consistent base line reference for further investigations
into ocean energetics, as well as for an intended inference calculation that is
more complete but also much more complex.
\end{abstract}

\section*{Plain Language Summary}
Energy plays an important role in quantifying the magnitude of motions at
different time and spatial scales. Many different dynamical processes contribute
to energy transfers within the ocean, and constraining the rate of transfer
remains a formidable challenge. This work provides a leading-order constraint on
the overall magnitude and spatial distribution of an eddy energy dissipation
time-scale, which relates to the rate of energy transfer out of the motions at
10 to 100 km in the ocean, where rotation and stratification play a leading
order role in the dynamics. A time-scale is ``backed out'' from a model via an
inverse approach: given a model for the eddy energy evolution and what we should
end up with (the eddy energy signature), what should we have started off with in
the first place (the dissipation time-scale)? Although our solution should be
interpreted as a lower bound given the assumptions going into the calculation,
it serves as an important physically consistent base line reference for further
investigations into ocean energetics, as well as for an intended inference
calculation that is more complete but also much more complex.

%% ------------------------------------------------------------------------ %%
%
%  TEXT
%
%% ------------------------------------------------------------------------ %%

%%% Suggested section heads:
% \section{Introduction}
%
% The main text should start with an introduction. Except for short
% manuscripts (such as comments and replies), the text should be divided
% into sections, each with its own heading.

% Headings should be sentence fragments and do not begin with a
% lowercase letter or number. Examples of good headings are:

% \section{Materials and Methods}
% Here is text on Materials and Methods.
%
% \subsection{A descriptive heading about methods}
% More about Methods.
%
% \section{Data} (Or section title might be a descriptive heading about data)
%
% \section{Results} (Or section title might be a descriptive heading about the
% results)
%
% \section{Conclusions}

%Text here ===>>>

%\cite<e.g.>[post note]{Marshall-et-al12}
%\citeA{Rathgeber-et-al17}

%%%%%%%%%%%%%%%%%%%%%%%%%%%%%%%%%%%%%%%%%%%%%%%%%%%%%%%%%%%%%%%%%%%%%%%%%%%%%%%%

\section{Introduction}\label{sec:intro}

The ocean, being a key component of the Earth system, plays a central role in
the Earth's energy and biogeochemical cycles through its ability to store and
transport large amounts of tracers \cite<e.g.,>{ZhangVallis13, Adkins13,
Ferrari-et-al14, Burke-et-al15, Bopp-et-al17, Jansen17, Takano-et-al18,
GalbraithdeLavergne19}. A key component is the transport provided by the ocean
circulation, taken here to mean both the large-scale mean circulation, as well
as the smaller-scale eddy motions. The large-scale mean motions tend to generate
smaller-scale eddy motions via instabilities, but the smaller-scale eddy motions
also interact and feedback onto the large-scale mean. The ability to represent
the multi-scale interaction faithfully (via explicit or sub-grid modeling
means), besides the theoretical interest, is of central importance to ocean
model performance, which impacts our ability to predict and assess impacts
within the ocean component as well as the wider Earth system
\cite<e.g.>{Hewitt-et-al17, FoxKemper-et-al19}.

An important quantity relating to dynamics at the relevant space and/or
time-scales is the energy, and there has been ongoing theoretical, numerical and
observational research into quantifying and constraining ocean energy content
and pathways. \citeA{FerrariWunsch09, FerrariWunsch10} provide one of the more
recent reviews of the global energy content and pathways of the overall ocean
energetic cycles, though the authors note that there are still relatively large
uncertainties associated with the magnitudes as well as spatial distributions in
the energy fluxes. From an ocean modeling point of view, energetically
constrained parameterizations have been increasingly proposed and investigated,
and such parameterizations have led to model improvements
\cite<e.g.>{Gaspar-et-al90, Cessi08, EdenGreatbatch08, MarshallAdcroft10,
Marshall-et-al12, OlbersEden13, Eden-et-al14, JansenHeld14, Mak-et-al18,
Nielsen-et-al18, Bachman19, Jansen-et-al19, Mak-et-al22}. Beyond improving our
understanding of the dynamics within the ocean, constraints for the energy
pathways play an important role in limiting the magnitude and form of energy
transfers between different dynamical components, which is also expected to lead
to improved performance of numerical ocean models via improvements to sub-grid
parameterizations of dynamical processes.

The focus of the present work is on providing a leading order constraint on the
energy pathway associated with the ocean geostrophic mesoscales, where the
dynamics are strongly constrained by rotation and stratification. Analogous to
processes such as internal waves where the impact of energy supply and removal
will affect quantities such as diapycnal mixing, the energy content and pathways
in the geostrophic mesoscale is expected to impact eddy induced advection and
isoneutral diffusion, usually quantified by an eddy induced velocity coefficient
or the Gent--McWilliams coefficient \cite{GentMcWilliams90, Gent-et-al95} and an
isoneutral diffusion coefficient \cite{Redi82, Griffies98}, respectively. Eddy
energy content and removal rates will affect the associated eddy--mean-flow
feedbacks, with consequences for the global overturning circulation and
stratification profile. Such an influence was demonstrated, for example, in the
work of \citeA{Mak-et-al22} for prognostic calculations within a global ocean
circulation model via an eddy energy constrained mesoscale eddy parmaeterization
scheme \cite<GEOMETRIC;>{Marshall-et-al12, Mak-et-al18}. In that work, the model
global ocean circulation and stratification were found to be acutely sensitive
to modest changes in an eddy energy dissipation time-scale $\lambda^{-1}$, and
were comparable sensitivities with respect to significant variations (halving
and doubling) in the Southern Ocean wind forcing magnitude. While we do not
expect such significant changes in the Southern Ocean wind forcing
\cite<e.g.>{Lin-et-al18}, analogous constraints for the eddy energy dissipation
time-scale, related in turn to the eddy energy flux out of the geostrophic
mesoscales, are lacking.

While the transfer of energy into the mesoscale is known to be primarily via
baroclinic instability, accessing the available potential energy at the large
planetary scales arising from large-scale wind and thermodynamic forcing
\cite{FerrariWunsch09, FerrariWunsch10}, there are many dynamical processes that
can lead to energy fluxes out of the mesoscale. These include but are not
limited to: direct return to the mean flow, via an inverse cascade in
rotationally dominant quasi-two-dimensional systems \cite<e.g.>{Salmon80,
Jansen-et-al19, Bachman19}; relative wind stress effects, whereby the
atmospheric wind forcing can spindown baroclinic eddies
\cite<e.g.>{Zhai-et-al12, Xu-et-al16, Rai-et-al21}; bottom drag
\cite<e.g.>{Sen-et-al08}; non-propagating form drag arising from bathymetric
form stress \cite<e.g.>{Klymak18, Klymak-et-al21}; scattering into lee waves by
geostrophic flow interaction with the bottom topography
\cite<e.g>{NikurashinFerrari11, Nikurashi-et-al13, Melet-et-al14, Melet-et-al15,
Yang-et-al18, Yang-et-al21}; loss of balance, from secondary fluid instabilities
of the mesoscale eddy motions themselves \cite<e.g.>{Molemaker-et-al05,
Barkan-et-al17, Rocha-et-al18, Chouksey-et-al18}. While there are suggestions
that some processes are more efficient at transferring energy out of the
mesoscale \cite<e.g., non-propagating form drag;>{Klymak-et-al21}, an overall
quantification on the geographical distribution of the energy transfer rate is
absent.

%, although we note
%the work of {\color{red}MarshallZhai}, who estimate an eddy residence time-scale
%(analogous to the dissipation time-scale in this work) from a ratio of the
%depth-integrated total (kinetic and potential) eddy energy and wind input of
%energy into the ocean. The former is estimated from satellite altimetry
%projected on the first baroclinic mode and ARGO float data
%\cite{Roullet-et-al14}, and the latter from atmospheric reanalyses and satellite
%altimetry. The procedure in {\color{red}MarshallZhai} requires large spatial
%averaging (zonal average and a relatively wide latitudinal window to obtain an
%estimate a latitudinally rolling average) resulting in a latitudinally varying
%energy residence time-scale (see Fig.~\ref{fig:lam_opt_eps5e15}$b$ in this work
%for their result).

Given the preceding discussion, we are primarily interested in providing an
overall estimate for the spatial distribution of the eddy energy dissipation
time-scale $\lambda^{-1}$, to serve as a leading order constraint for (1) the
energy fluxes out of the geostrophic mesoscales, (2) the various dynamical
components leading to the aforementioned energy flux, (3) the development of
further parameterizations of the energy fluxes, and (4) further theoretical,
numerical and observation works into the ocean energetic cycles. In this work,
we perform an inference calculation for a spatially varying $\lambda^{-1}$ via
an inverse approach, and illustrate the approach via the GEOMETRIC description
for the geostrophic mesoscale eddy energy evolution \cite<although other choices
are possible, e.g. MEKE from>{Jansen-et-al19}. Such inference calculations have
a beneficial side effect of providing a way to tune for uncertain parameters for
use in prognostic numerical models.

As will be detailed and argued later, while one would ideally like to carry out
a \emph{dynamically} constrained inference calculation, the associated
calculation is numerically complex and computationally intensive \cite<e.g.
requiring derivation of adjoint models given the expected large amount of
degrees of freedom to maintain computational feasibility;>{Wunsch-inverse,
Kalnay-DA, Gunzburger-control}. As a precursor to the proposed difficult but
more complete dynamically constrained inference calculation, we report here a
kinematic/diagnostic-type inference calculation that is much simpler and
computationally inexpensive, with the caveat that there are simplifying
assumptions involved, resulting in a spatially varying $\lambda^{-1}$ that will
be argued to be a physically plausible \emph{lower bound}. In that sense, the
present work reports a calculation that can serve as a prior the proposed
dynamically constrained inference calculation. Additionally, the present work
provides some exposure to the methodology and associated computational tools
that are perhaps less well-known in the field of earth system science.

In \S\ref{sec:method} we provide details to the mesoscale version of the
GEOMETRIC parameterization, an overview of the inference and inverse approach,
and the implementation details of the diagnostic-type inference problem for the
geostrophic mesoscale eddy energy dissipation time-scale $\lambda^{-1}$.
Analyses of the characteristics of the resulting solutions are detailed in
\S\ref{sec:analysis}. In \S\ref{sec:prognostic} we investigate the consequences
of utilizing the inferred solution in a prognostic model, providing some support
that the reported dissipation time-scale is a physically plausible lower bound.
In \S\ref{sec:conclusion} we provide outlooks on the proposed dynamically
constrained inference calculation to improve our understanding of ocean energy
pathways, parameterization of energy pathways, and for inferring uncertain
system parameters.

%-------------------------------------------------------------------------------

\section{Methodology and implementation}\label{sec:method}

%%%%%%%%%%%%%%%%%%%%%%%%%%%%%%%

\subsection{GEOMETRIC}

For this work, we consider the GEOMETRIC description \cite{Marshall-et-al12} as
a model for the evolution of mesoscale eddy energy. The choice of GEOMETRIC is
made for its theoretical foundations \cite{Marshall-et-al12, MaddisonMarshall13}
and its demonstrated capabilities in recovering some key ocean sensitivities
possessed by high resolution numerical models that permit and/or resolve
mesoscale eddies in prognostic coarse resolution models \cite{Mak-et-al18,
Mak-et-al22}; however we stress again that other choices such as MEKE from
\citeA{Jansen-et-al19} would be possible with suitable adaptations of the
methodology detailed below. Denoting $\hat{E} = \int E\; \mathrm{d}z$ as the
depth-integrated total eddy energy, the mesoscale version of GEOMETRIC suggests
we take \cite<cf. Eq. 1 of>{Mak-et-al22}
\begin{equation}\label{eq:GEOM-gm}
  \kappa_{\rm gm} = \alpha\frac{\hat{E}}{\int M^2 / N\; \mathrm{d}z},
\end{equation}
where $\kappa_{\rm gm}$ is the eddy induced velocity coefficient
\cite<c.f.>{GentMcWilliams90, Gent-et-al95}, $\alpha$ is a non-dimensional
tuning parameter, and $M^2 = |\nabla_H b|$ and $N^2 = \partial b / \partial z$
are the mean horizontal and vertical buoyancy gradients respectively. In a
prognostic calculation, $\hat{E}$ evolves according to the parameterized eddy
energy budget \cite<cf. Eq. 2 of>{Mak-et-al22}
\begin{equation}\label{eq:GEOMloc-e}
  \frac{\mathrm{d}\hat{E}}{\mathrm{d}t}
    + \underbrace{\nabla_H \cdot \left( \left(\widetilde{\mathbf{u}}^z - |c|\, \mathbf{e}_x\right) \hat{E} \right)}_\textnormal{advection}
   = \underbrace{\int \kappa_{\rm gm} \frac{M^4}{N^2}\; \mathrm{d}z}_\textnormal{source}
    - \underbrace{\lambda \hat{E}}_\textnormal{dissipation} 
    + \underbrace{\eta_E\nabla^2_H  \hat{E}}_\textnormal{diffusion},
    %\tag{GEOMloc-e}
\end{equation}
where $\widetilde{\mathbf{u}}^z$ is the depth-averaged mean flow, $|c|$ is the
long Rossby wave phase speed, $\lambda$ is a linear eddy energy dissipation rate
(so that $\lambda^{-1}$ is the eddy energy dissipation time-scale of principal
focus in this work), and $\eta_E$ is an eddy energy diffusion coefficient. From
prognostic calculations in \citeA{Mak-et-al22}, the dominant contributions in
Eq.~(\ref{eq:GEOMloc-e}) were found to be from the source and dissipation, with
secondary contributions from diffusion and advection, though the latter two
terms are important for the resulting eddy energy spatial distribution.

The principal aim of this work will be on constraining the spatial distribution
of $\lambda^{-1}(\phi, \theta)$ (where $\phi$ and $\theta$ denote the longitude
and latitude respectively), here taken to encapsulate all the aforementioned
dynamical processes that leads to an energy flux out of the mesoscale. This is
of course a rather drastic approximation, although there is some suggestion that
a dominant source of eddy energy removal from the mesoscale could be from
non-propagating form drag \cite{Klymak18, Klymak-et-al21}, and the energy
transfer arising from non-propagating form drag is better represented as a
linear drag and expected to largely depend on the bathymetry. We also make the
simplifying assumptions of taking both $\eta_E$ and $\alpha$ to be a prescribed
constant. For $\eta_E$, this is partially justified \emph{a priori}, where we
expect the diffusion term to be of secondary importance in the evolution of the
parameterized eddy energy in prognostic calculations, and \emph{a posteriori},
where the inferred solutions were found not to be overly sensitive to the choice
of $\eta_E$. For $\alpha$, this is mainly for simplicity, and it is known
$\alpha$ does vary in space \cite{Poulsen-et-al19}. The inferred solution will
be seen to display some sensitivity to the value of $\alpha$ through its role in
the source term. The possibility of a joint inference calculation for
$\lambda^{-1}$ and $\alpha$ are discussed in \S\ref{sec:conclusion}.

%%%%%%%%%%%%%%%%%%%%%%%%%%%%%%%

\subsection{Parameter inference problem}

The inference calculation for $\lambda^{-1}(\phi, \theta)$ here utilizes the
variational approach \cite<e.g.>{Kalnay-DA}. In the general case, we have state
variables $\mathbf{w}$ that depend on control variable $\lambda^{-1}(\phi,
\theta)$ via some model $F(\mathbf{w}; \lambda^{-1}) = 0$, and the aim is to
seek $\lambda^{-1}(\phi, \theta)$ such that the mismatch between some target
data $\mathbf{w}_{\rm data}$ and $\mathbf{w}$ is minimized, possibly subject to
some regularization $\mathcal{R}(\mathbf{w}; \lambda)$ tht encapsulates our
prior expectations for the state and/or control variables. A variational
approach is for example utilized in the Estimating the Circulation and Climate
of the Ocean framework \cite<ECCO; e.g.>{Forget-et-al15a, Fukumori-et-al18}: the
ocean state variables (e.g. temperature, salinity) are the state variables in
ECCO, and the ECCO framework seeks to adjust the control variables (e.g. wind
forcing, initial ocean state, parameterization parameters) with the aim to
minimize the mismatch to observational data (e.g. sea surface height,
hydrographic sections, currents strengths) over time, subject to regularizations
(e.g. the wind forcing not deviating too far from climatology) and the
constraint that the calculated state satisfies the dynamical equations as
implemented in MITgcm \cite{Marshall-et-al97a, Marshall-et-al97b}. Within ECCO,
the optimization problem is achieved through an adjoint method
\cite<e.g.,>{Wunsch-inverse, Kalnay-DA, Gunzburger-control}.

Ultimately the aim would be to carry out a dynamically constrained inference
problem, where the parameterized eddy energy equation Eq.~(\ref{eq:GEOMloc-e})
would be coupled to an ocean global general circulation model, so that the
evolution of the parameterized eddy energy profile is dynamically and
self-consistently interacting with the evolving the ocean state (e.g. $M^2$,
$N^2$, $\widetilde{\mathbf{u}}^z$ and $|c|$ etc.), and infer for the spatially
varying $\lambda^{-1}(\phi, \theta)$ or other choices of control variables. Such
an inference calculation could leverage the ECCO framework with appropriate
modifications of the constraining equations. However, such an endeavor will
require a significant investment in human development time and computational
resources, exacerbated by the fact that we have no leading order constraints of
$\lambda^{-1}(\phi, \theta)$ to serve as a prior for regularizing the inverse
problem. Thus, with the dynamically constrained inference as the eventual goal,
in this work we present a useful complementary calculation that serves to
provide a first leading order estimate for $\lambda^{-1}(\phi, \theta)$. We note
that we can in principle diagnose the mean state variables required for the eddy
energy equation Eq.~(\ref{eq:GEOMloc-e}) from a high resolution global
circulation model, and supply the inference calculation with prescribed physical
state variables. The result is that we substantially reduce the complexity of
the constraining model $F(\mathbf{w}; \lambda^{-1}) = 0$, since we no longer
need the calculation to be coupled to a dynamical model. A drawback however is
that dynamical feedbacks are removed; and consequences are discussed at the end
of the present subsection.

With sufficiently long time-averaging and with the assumption that the evolution
of the eddy energy has reached a statistically stead state, the constraining
model becomes an elliptic problem given by
\begin{equation}\label{eq:model}
   F(\hat{E}; \lambda^{-1}) \equiv \underbrace{\eta_E\nabla^2_H  \hat{E}}_\textnormal{diffusion} - \underbrace{\lambda \hat{E}}_\textnormal{dissipation} + \underbrace{\alpha \frac{\int M^4/N^2\; \mathrm{d}z}{\int M^2 / N\; \mathrm{d}z} \hat{E}}_\textnormal{source} - \underbrace{\nabla_H \cdot \left( \left(\widetilde{\mathbf{u}}^z - |c|\, \mathbf{e}_x\right) \hat{E} \right)}_\textnormal{advection} = 0,
\end{equation}
where we have substituted for the GEOMETRIC prescription of $\kappa_{\rm gm}$
given by Eq.~(\ref{eq:GEOM-gm}) into the eddy energy source term. The control
variable for the present problem will be $\lambda^{-1}(\phi,\theta)$, and the
state variable will be $\hat{E}(\phi, \theta)$. With a constraining model such
as Eq.~(\ref{eq:model}), we seek the optimal $\lambda^{-1}(\phi,\theta)$ that
minimizes the cost functional
\begin{equation}\label{eq:cost}
  J \equiv J_1 + J_2 = \frac{1}{A}\left\|\hat{E} - \hat{E}_{\mathrm{data}}\right\|^2_{L^2} + \epsilon \|\nabla_H \lambda^{-1}\|^2_{L^2} \qquad \left(\| f \|^2_{L^2} = \int_A f^2\; \mathrm{d}A\right),
\end{equation}
where $A$ is the two-dimensional domain and $\mathrm{d}A$ is the area element.
The optimization problem seeks for $\hat{E}(\lambda^{-1})$ that is close to the
diagnosed $\hat{E}_{\rm data}$ in the $L^2$ norm (the spatial integral of the
square of the mismatches), measured here by the cost functional $J_1$.
Generally, optimization problems without a regularization term are ill-posed,
leading to numerical non-convergence. For this work, we employ a Tikanov-type
regularization, introduced via a penalization term on the gradients of the
control variable, given here by the cost functional $J_2$, with strength measure
by some parameter $\epsilon$. Given that we are employing high resolution global
circulation model data to infer for $\lambda^{-1}(\phi, \theta)$ (which is a
representation of the processes on a coarse resolution grid), the penalization
term could be thought as our prior belief that our control variable should have
broad spatial structures, or as a coarse-graining or averaging operation for the
control variable $\lambda^{-1}(\phi,\theta)$. If $\epsilon$ is too small, the
optimization problem is trying to match per grid point, leading to extreme
values in $\lambda^{-1}(\phi,\theta)$ and sometimes numerical non-convergence.

Before we move on, we caveat that, by prescribing the dynamical variables, the
constraining equation Eq.~(\ref{eq:model}) is formally linear in the state
variable $\hat{E}$. The resulting inference calculation then becomes relatively
easy to implement and computationally inexpensive to perform, since this is
essentially an optimization calculation subject to an elliptic partial
differential equation under the present kinematic/diagnostic approximation.
However, in a fully dynamical setting, the isopycnal slopes $M^2/N^2$ for
example would be regarded as an implicit function of $\kappa_{\rm gm}$, which
itself would be an explicit function of $\hat{E}$ through the mesoscale eddy
parameterization. The growth of $\hat{E}$ will lead to an increase in
$\kappa_{\rm gm}$, which would lead to a flattening of the isopycnals, i.e.
reductions in $M^2$, and the growth rate of the eddy energy would thus normally
be self-limiting. Without dynamical feedbacks, the growth rate is expected to be
over-estimated. Since we expect the dominant balance in the eddy energy equation
Eq.~(\ref{eq:GEOMloc-e}) to be between the source and dissipation term, for a
fixed target $\hat{E}_{\mathrm{data}}$, the present inference procedure is
expected to return a $\lambda^{-1}(\phi, \theta)$ with values that are too small
(i.e., a dissipation time-scale that is too short). The results presented here
should thus be viewed as a \emph{lower bound} for $\lambda^{-1}(\phi,\theta)$.
Some additional evidence in support of the proposed interpretation of the
solutions as a lower bound is given in via prognostic calculations with full
dynamical feedbacks in \S\ref{sec:prognostic}.

%%%%%%%%%%%%%%%%%%%%%%%%%%%%%%%

\subsection{Forcing and target data diagnoses for inference calculation}

For the proposed inference calculation, the forcing data we require are
depth-integrated values of $M^2$, $N^2$, $\widetilde{\mathbf{u}}^z$, $|c|$, and
the target $\hat{E}_{\rm data}$. For this work, all the aforementioned variables
were diagnosed from the nominally $1/12^\circ$ horizontal resolution Nucleus for
European Modelling of the Ocean \cite<NEMO;>{Madec-NEMO} ORCA0083-N01 hindcast
outputs (see Data Availability section). Data used here were calculated from the
five-day averaged outputs between and inclusive of the simulation years 2006 to
2010 (the last five years of the ORCA0083-N01 calculation), as depth-integrals
on the native NEMO tri-polar grid \cite{MadecImbard96}, and then time-averaged
over the five year period.

The depth-integrated total eddy energy $\hat{E}$ is computed as the sum of the
depth-integrated eddy kinetic and potential energy. The depth-integrated
(specific) eddy kinetic energy (with units of $\mathrm{m}^3\ \mathrm{s}^{-2}$)
is defined in the usual fashion as
\begin{equation}
  \mbox{EKE} = \frac{1}{2}\int_{-H}^0 \left(\overline{\mathbf{u}\cdot\mathbf{u}} - \overline{\mathbf{u}}\cdot\overline{\mathbf{u}}\right)\; \mathrm{d}z,
\end{equation}
where $\overline{(\cdot)}$ is a time average. To obtain the analogous
depth-integrated eddy potential energy, we take the five-day averaged
temperature and salinity outputs, convert into neutral density $\gamma^a$
co-ordinates via the \citeA{McDougallJackett05} expression, re-bin the
variables into an interval $[\gamma^a_t = 1020\ \mathrm{kg}\ \mathrm{m}^{-3},
\gamma^a_b = 1029\ \mathrm{kg}\ \mathrm{m}^{-3}]$ (with smaller bin widths
towards $\gamma^a_b$), calculate the depth associated with the neutral density
$z(\gamma^a)$ every five days over the five year period, and compute
\begin{equation}\label{eq:epe}
  \mbox{EPE} = \frac{1}{2}\int_{\gamma^a_b}^{\gamma^a_t} g \left(\overline{z^2} - \overline{z}^2\right)\; \mathrm{d}\gamma^a.
\end{equation}

For the stratification parameters $M^2 / N$ and $M^4 / N^2$ as ratios, first we
compute $N^2$ and $N$ as a three-dimensional field using the CDFTOOLS package
from five-day averaged temperature and salinity data (using \texttt{cdfbn2}; see
Data Availability section). A new program (\texttt{cdfsn2}) was created to
mirror the NEMO computation of stratification gradient parameters, with code
largely taken from the NEMO \texttt{ldfslp} subroutine and dependencies. The new
\texttt{cdfsn2} routine computes the isopycnal slopes $M^2 / N^2$ (\texttt{wslp}
in NEMO), appropriately modified by slope limiters, partial step corrections,
and tapering as the lateral, mixed layer and surface boundaries are approached,
and passed through a Shapiro filter in the horizontal. The intermediary $N$ and
$N^2$ are then multiplied to the slope variables to obtain $M^4/N^2$ and $M^2/N$
as three-dimensional fields, which are then depth-integrated and time-averaged.
The \texttt{cdfsn2} routine is able to reproduce the NEMO outputs directly up to
very minor discrepancies (code tested on a simple re-entrant channel model and
outputting the \texttt{wslp} variable and its modifications directly).

The depth-averaged mean flow $\widetilde{\mathbf{u}}^z$ is computed in the usual
way from the available data. The long Rossby phase speed $|c|$ is computed as
\cite<e.g.,>[Eq.~12.3.13]{Gill-GFD}
\begin{equation}
  |c| = \frac{c_1^2 \cos\theta}{2\Omega R_a \sin^2 \theta}, \qquad c_1 = \frac{1}{\pi}\int_{-H}^0 |N|\; \mathrm{d}z,
\end{equation}
where $\theta$ is the latitude, $\Omega$ is the Earth's angular frequency, $R_a$
is the Earth's radius, and $c_1$ is an approximation of the first baroclinic
phase speed \cite<e.g.>{NurserBacon14}. This phase speed is computed in this
case from the time-averaged $\int |N|\; \mathrm{d}z$ field directly. One thing
to note is that, because of the NEMO tri-polar grid, the ($u,v$) components of
the velocity only correspond exactly to the zonal and meridional velocities when
south of around $20^\circ\ \textnormal{N}$ \cite{MadecImbard96}. While the
anisotropy is relatively small, there are minor inconsistencies when
interpolating data to and from the tri-polar grid. The effect of this error is
not expected to significant, since the advective contributions are expected to
be rather minimal (and is supported by analyses presented in
\S\ref{sec:analysis}).

All the high resolution processed data on the tri-polar grid were additionally
passed through a diffusion-based filter \cite{Grooms-et-al21} to smooth out
variations smaller than twelve grid points, i.e. filtering to obtain the fields
that are coherent over a nominally $1^\circ$ horizontal resolution. This
filtering step does not appear to be strictly necessary for the conclusions
presented here, since the time-averaging operation already removes a substantial
portion of the fluctuations below the $1^\circ$ horizontal resolution for the
variables of interest here \cite<cf.>{Rai-et-al21}. Sample calculations without
the filtering step led to no noticeable changes in the results presented.

%%%%%%%%%%%%%%%%%%%%%%%%%%%%%%%

\subsection{Numerical implementation of inference problem}

For the implementation of the inference calculation, we leverage the Firedrake
software \cite{Rathgeber-et-al17}, an automatic code generation framework with
high level specification in Python that utilizes the finite element formalism
\cite<e.g.,>[Ch.6]{Durran-Numerical}. The procedure is that, given a finite
element mesh, we specify the function space on which we seek our solution, taken
to be continuous Galerkin with first-order Lagrange polynomials as the basis
here. We then implement our constraining model in the weak form
\cite<e.g.,>[Ch.1]{Evans-PDE}, which in this case involves multiply by a test
function with sufficient differentiability such that integration by parts maybe
performed, and implement any natural or imposed boundaries on the problem
accordingly. For a scalar test function $\psi$, the weak form
$\mathcal{F}(\hat{E}; \lambda) = 0$ associated with the constraining model
Eq.~(\ref{eq:model}) is given by
\begin{eqnarray}\label{eq:model-weak}
   \mathcal{F}(\hat{E}; \lambda) &\equiv& \int_A \left[\eta_E\nabla_H \hat{E} \cdot\nabla_H \psi + \left(\lambda \hat{E} - \alpha \frac{\int M^4/N^2\; \mathrm{d}z}{\int M^2 / N\; \mathrm{d}z} \hat{E}\right)\psi \right.\nonumber \\
   &&  \hspace{5mm} \left. - \left( \left(\widetilde{\mathbf{u}}^z - |c|\, \mathbf{e}_x\right) \hat{E} \cdot \nabla_H \psi \right)\right]\; \mathrm{d}A = 0.
\end{eqnarray}
The boundary terms arising from integration by parts are identically zero from
the boundary conditions for this problem, namely, $\nabla_H \hat{E}\ \cdot\
\mathbf{n} = 0$ and no-normal flow conditions $\left(\widetilde{\mathbf{u}}^z -
|c|\, \mathbf{e}_x\right)\ \cdot\ \mathbf{n} = 0$, where $\mathbf{n}$ is the
vector perpendicular to the domain boundary $\partial A$. As opposed to the more
conventional strong form formulation, where we seek a solution that satisfies
Eq.~(\ref{eq:model}) point-wise and the solution $\hat{E}$ is required to be
twice differentiable, the weak form formalism only requires a weak form solution
$\hat{E}$ to satisfy Eq.~(\ref{eq:model-weak}) in an integral sense, has weaker
assumptions on differentiability, and the resulting problem is readily solved
numerically within Firedrake.

Since Firedrake employs the finite element framework, we need a finite element
mesh. For this work a choice was made to solve the problem on a two-dimensional
spherical mesh embedded into the standard three-dimensional Euclidean space
$\mathbb{R}^3$, where the relevant periodicities and land boundaries are built
into the mesh itself (as opposed to constructing a longitude-latitude grid with
North Pole folding, which leads to a point singularity). An unstructured mesh
with triangular elements with characteristic length-scale $100$ km was created
using the \texttt{Qmesh} package \cite{Avdis-et-al18}, with the land shape
generated from the NEMO ORCA1 configuration (a global ocean configuration with a
nominal horizontal resolution of $1^\circ$). To move the diagnosed data from the
tri-polar NEMO grid onto the mesh, a simple change of co-ordinates from ($\phi,
\theta$) to ($x,y,z$) and linear interpolation (through the Python
\texttt{scipy} package) was performed, with extrapolations where necessary (e.g.
near the geographical locations associated with the edges of the tri-polar
grid). Note that while interpolating scalars is straight-forward, interpolating
vectors require preserving both the direction and magnitude, which can be
achieved via a multiplication by a rotation matrix.

Once we have data on the mesh, and upon specification of the parameters, we can
proceed to build the constraint model, couple the model to an optimizer in
Firedrake, and solve the resulting optimization problem. For numerical stability
reasons, in order to obtain physical solutions, we modify
Eq.~(\ref{eq:model-weak}) by adding a term $\hat{E}_0$ that maintains a minimum
energy value, and replacing $\hat{E}$ by $\hat{E}_{\rm data}$ in the source term
so that the source term becomes a diagnostic variable, i.e. we solve
\begin{eqnarray}\label{eq:model-weak1}
   \mathcal{F}(\hat{E}; \lambda) &\equiv& \int_A \left[\eta_E\nabla_H \hat{E} \cdot\nabla_H \psi + \left(\lambda (\hat{E} - \hat{E}_0) - \alpha \frac{\int M^4/N^2\; \mathrm{d}z}{\int M^2 / N\; \mathrm{d}z} \hat{E}_{\mathrm{data}}\right)\psi \right.\nonumber \\
   &&  \hspace{5mm} \left. - \left( \left(\widetilde{\mathbf{u}}^z - |c|\, \mathbf{e}_x\right) \hat{E} \cdot \nabla_H \psi \right)\right]\; \mathrm{d}A.
\end{eqnarray}
For the first modification, adding a $\hat{E}_0 > 0$ term is necessary for the
numerical solver to converge to a non-zero solution in the absence of the second
modification, otherwise its role is to ensure the solution $\hat{E}$ has a
minimum background value, consistent with how GEOMETRIC is currently implemented
into NEMO \cite{Mak-et-al22}. The second modification is perhaps a more severe
one, and arises most likely because we are forcing the present inference problem
with prescribed data and removing dynamical feedbacks by construction. With full
dynamical coupling, the stratification responds to changing $\hat{E}$ via the
GEOMETRIC parameterization, and is a self-limiting process that arrests the
growth of $\hat{E}$. In the present setup, this is not possible since the
dynamical variables are prescribed, and the form of the growth term implies
exponential growth of $\hat{E}$ during the iterations (through time-stepping or
an iterative solver). Any imbalance in the initialization of the data will be
amplified exponentially during the iterations. The imbalance will generically
occur in the present set up, since the data diagnosed from ORCA0083-N01 will
certainly not be in exact steady state, and we do not have the perfectly matched
$\lambda^{-1}(\phi, \theta)$ and $\hat{E}$ at the initialization stage, since
that necessarily implies we have the solution before even solving the problem.
Solving Eq.~(\ref{eq:model-weak}) as is leads to significant over/undershoots,
with very large positive and negative values in $\hat{E}$; during the
development of the present work, a channel model simulation could be integrated
to equilibrium, and the described problem was lessened though not completely
circumvented. Replacing $\hat{E}$ by $\hat{E}_{\rm data}$ in
Eq.~(\ref{eq:model-weak1}) acts as a numerical stabilizer that limits the amount
of growth possible. The consequences of the approximation and how to alleviate
it will be discussed in \S\ref{sec:conclusion}.

To solve the optimization problem where we minimize the cost functional
Eq.~(\ref{eq:cost}) subject to the model given by Eq.~(\ref{eq:model-weak1}), we
employ the \texttt{tlm\_adjoint} library \cite{Maddison-et-al19} with the
Firedrake wrapper, which allows us to build the constrained optimization problem
in Firedrake. The optimization problem is solved using a L-BFGS (Limited memory
Broyden--Fletcher--Goldfarb--Shanno) algorithm, which is a quasi-Newton method
that looks for descents through an estimate of the inverse Hessian matrix
\cite<e.g.,>{Byrd-et-al95}. As demonstration, the actual production code used
for implementing the elliptic solve for Eq.~(\ref{eq:model-weak1}), forming the
cost functional Eq.~(\ref{eq:cost}) and wrapping to the optimizer is given in
Fig.~\ref{fig:code}. Aside from the fact this code is incredibly short and
required minimal time to write, the optimization and inference calculations
become flexible since any changes in the problem statement are automatically
propagated to the optimizer routines via automatic code generation capabilities.
For example, if we want to modify the cost functional (e.g. changing the measure
of the mismatch, changing the existing Tikanhov-type regularization, adding
extra penalization terms), then we simply modify the relevant lines defining
$J$. If we want to modify the constraining model (e.g. remove advection, employ
quadratic dissipation, spatially varying $\alpha$, increasing the number of
control variables, use another choice of eddy energy equation), we re-define the
weak form \verb|F| and/or the control variables accordingly, as long as if we
have relevant diagnostic inputs to force the resulting equations.

\begin{figure}
  \begin{lstlisting}
    def forward(lam):
      Edz = Function(P, name = "Edz")
      F = ( + Constant(alp) * Etot_zint * N_over_M2 * M4_over_N2 * test * dx  # source
            - lam * ( Edz - Constant(E0) ) * test * dx                        # dissipation
            - Constant(nu)  * dot( grad(Edz), grad(test) ) * dx               # diffusion
            + dot(Edz * u_zavg, grad(test)) * dx                             # advection
           )
          
      solve(F == 0, Edz, solver_parameters = sp) # magic inherited from Patrick F.
      J = Functional(name="J") # cost functional: mismatch + regularisation
      J1 = (1.0 / domain_area) * inner(Edz - Etot_zint, Edz - Etot_zint) * dx
      J2 = Constant(eps) * inner(grad(lam), grad(lam)) * dx
      J.assign(J1 + J2)
      return Edz, J

    lam = Function(P, name = "lam"); lam.assign(Constant(lam_s))      # initial guess
    start_manager(); _, J = forward(lam); stop_manager() # start tape
    def forward_J(lam):              # define function to optimise: J = forward(lam)
        return forward(lam)[1]
    lam_opt, result = minimize_scipy(forward_J, lam,
                                         method="L-BFGS-B",
                                         options={"disp": True, "maxiter": 100})
  \end{lstlisting}
  \caption{Firedrake driver code for solving the optimization problem that
  minimizes the cost functional Eq.~(\ref{eq:cost}) subject to
  Eq.~(\ref{eq:model-weak1}). The code defines the weak form, solves it with
  some specified solver parameters, and defines the cost functional. The
  \texttt{tlm\_adjoint} equation manager is then called, and passes the defined
  cost functional $J = \mathcal{F}(\lambda^{-1})$ to the optimization
  algorithm.}
  \label{fig:code}
\end{figure}

As a final note, while the inference calculation is carried out on the spherical
mesh, it is desirable to transform data onto a regular longitude-latitude grid
for further analyses, and necessary to put data on NEMO ORCA1 tri-polar grid to
test out consequences of employing the inferred $\lambda^{-1}(\phi, \theta)$ in
prognostic calculations. A technical complication arises here that the spherical
mesh is an immersed manifold (a sub-manifold of the 2-sphere $S^2$ embedded in
$\mathbb{R}^3$), which has zero measure in the embedding space ($S^2$ has no
`volume'). The standard procedure of probing the function values on the immersed
manifold will fail as it is mathematically ill-defined: the probability of a
given co-ordinate point being on the manifold is related to the measure, and an
arbitrary co-ordinate point we wish to query the function value at will almost
surely not be on the zero-measure manifold. One way round this technical issue
is to probe the finite element output using the \texttt{vtk} package directly,
locating the cell closest to the co-ordinate of the point being queried, project
that point onto the cell, which is now regarded as a sub-manifold of
$\mathbb{R}^2$ (the measure of interest is now `area'), and the function can
then be queried accordingly. For the analyses presented in \S\ref{sec:analysis},
all relevant mesh data was interpolated onto a regular latitude-longitude grid
with $1/4^\circ$ spacing for ease of visualization. For the prognostic
calculations in \S\ref{sec:prognostic}, the inferred $\lambda^{-1}(\phi,
\theta)$ was interpolated directly from the spherical mesh onto the NEMO ORCA1
tri-polar grid.

%-------------------------------------------------------------------------------

\section{Analysis of inferred dissipation time-scale $\lambda^{-1}(\phi, \theta)$}\label{sec:analysis}

Inference calculations for $\lambda^{-1}(\phi, \theta)$ with the cost functional
Eq.~(\ref{eq:cost}) subject the constraint Eq.~(\ref{eq:model-weak1}) were
performed using the parameters in Table~\ref{tbn:param}, with values chosen
partly to coincide with those used in the prognostic calculations documented in
\citeA{Mak-et-al22}. All optimization calculations were initialized with a
spatially uniform initial condition of $\lambda_0^{-1} = 365$ days and $\hat{E}
= 0$. Of principal interest here is the control calculation utilizing $\epsilon
= 5\times10^{15}$.

\begin{table}
  \caption{Parameter values employed in the inference problem. The boldfaced
  values denote the control calculation.}
  \label{tbn:param}
  \centering
  \begin{tabular}{l c c}
  \hline
  Parameter  & Values & Units\\
  \hline
   $\alpha$         & 0.04  & ---\\
   $\eta_E$         & 500   & $\mathrm{m}^2\ \mathrm{s}^{-1}$\\
   $\hat{E}_0$      & 4.0   & $\mathrm{m}^3\ \mathrm{s}^{-2}$\\
   $\lambda_0^{-1}$ & 365   & days\\
   $\epsilon$       & $a \times 10^b$ & $\mathrm{m}^8\ \mathrm{s}^{-2}$\\
                    & $a \in \{1,\mathbf{5}\}$, $b \in \{13, 14, \mathbf{15}, 16\}$ &\\
  \hline
  \multicolumn{3}{l}{}
  \end{tabular}
\end{table}

\begin{figure}
  \includegraphics[width=\textwidth]{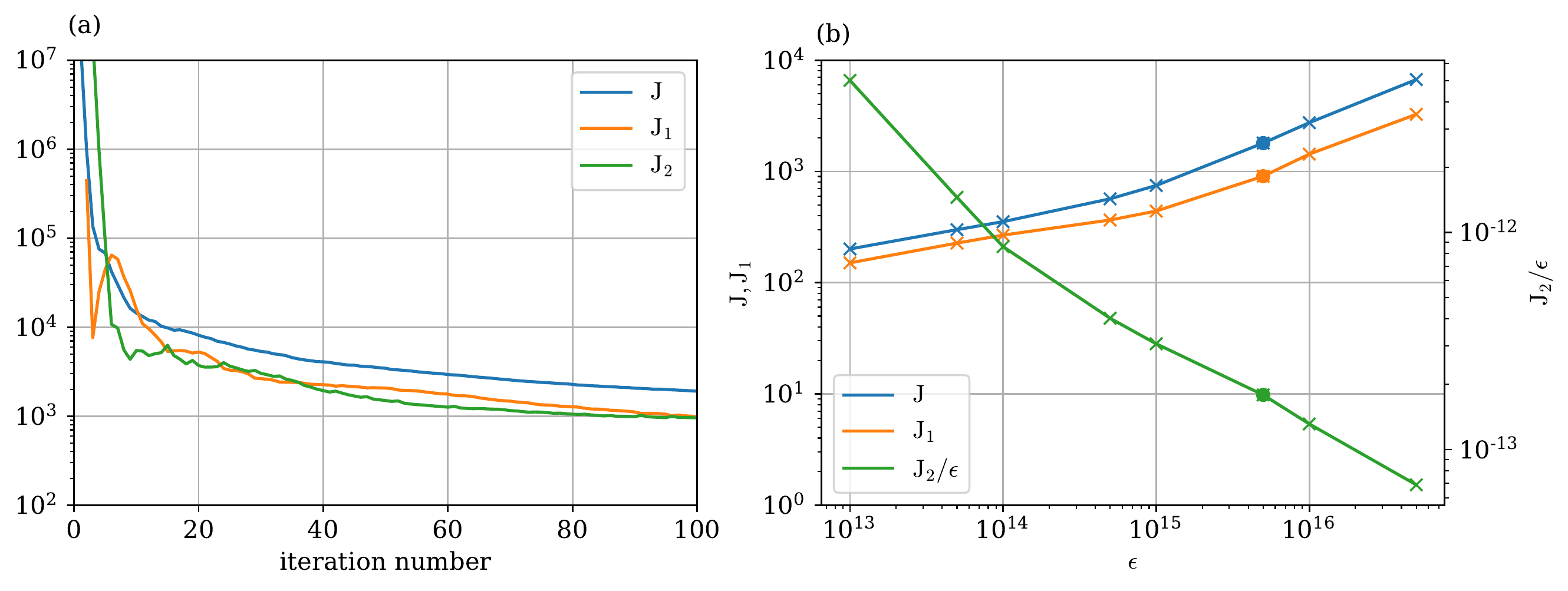}
  \caption{Behavior of the cost functional Eq.~(\ref{eq:cost}) as a function of
  ($a$) iterations for the $\epsilon=5\times10^{15}$ calculation, and ($b$)
  as a function of $\epsilon$. The markers denote the parameters with
  calculations, and the solid circle marker denotes the control calculation.}
  \label{fig:J_plots}
\end{figure}

For the optimization calculations, the default convergence criterion based on
projected gradients or the differences in function values were not triggered,
though it is clear that the values of the cost functional are converging to some
asymptotic value (albeit rather slowly); see for example
Fig.~\ref{fig:J_plots}$a$ for the control calculation as a function of the
iteration number. For consistency reasons, all results relating to the
optimization calculation reported in this work were taken to be the solution at
the 100${}^{\mathrm{th}}$ iteration. The corresponding run time for each
calculation was around three minutes on a laptop (the one utilized for this work
has Intel i7 CPUs and a RAM capacity of 8 GB). Fig.~\ref{fig:J_plots}$b$ shows
the dependence of the cost functional $J$ and its components on the value of
$\epsilon$. As expected, with increasing $\epsilon$, gradients in the optimized
solution, proportional to the numerical value of $J_2$, are penalized, at the
expense of having larger mismatches between the target $\hat{E}_{\rm data}$ and
the state variable $\hat{E}(\lambda^{-1})$, encoded by $J_1$.

\begin{figure}
  \begin{center}
  \includegraphics[width=\textwidth]{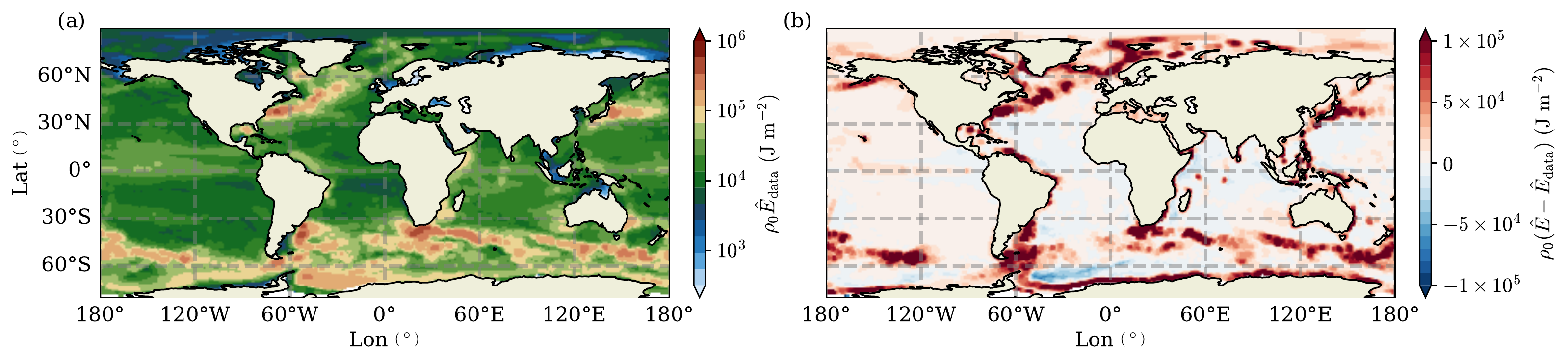}
  \end{center}
  \caption{($a$) The diagnosed depth-integrated eddy energy target $\rho_0
  \hat{E}_{\rm data}$ and ($b$) the signed mismatch $\rho_0 (\hat{E} -
  \hat{E}_{\rm data})$ associated with $\lambda^{-1}(\phi,\theta)$ with
  $\epsilon = 5\times 10^{15}$. Here we take $\rho_0=1026\ \mathrm{kg}\
  \mathrm{m}^{-3}$, and the variables are in units of $\mathrm{J}\
  \mathrm{m}^{-2}$.}
  \label{fig:Eerr_eps5e15}
\end{figure}

Fig.~\ref{fig:Eerr_eps5e15} shows the target $\hat{E}_{\rm data}$ and the signed
mismatch $\hat{E} - \hat{E}_{\rm data}$ of the control calculation as a function
of longitude and latitude. The target $\hat{E}_{\rm data}$ is that diagnosed
from the eddy resolving calculation ORCA0083-N01 \cite<cf. Fig.~5$a$
of>{Mak-et-al22}, showing a large eddy energy signature in the Southern Ocean
and Western Boundary Currents, and lower eddy activity in the Arctic and the
ocean basins. From the signed mismatch, the eddy energy signature associated
with the inferred solution is generally larger than the target. If $\epsilon$ is
decreased in magnitude, the local signed mismatches shown in
Fig.~\ref{fig:Eerr_eps5e15}$b$ decreases in magnitude, with negligible changes
in the position; however, that results in larger gradients and more extreme
values in $\lambda^{-1}(\phi,\theta)$, with the local signed mismatches still
largely skewed towards positive values. Recalling from \S\ref{sec:method}, the
stratification is prescribed in this work, thus forbidding dynamical
adjustments, and the positive skew observed here indicates that the now
prescribed eddy energy growth rate is still too large, requiring the
$\lambda^{-1}(\phi,\theta)$ to take smaller values to increase the dissipation
to balance the growth rate. As argued in \S\ref{sec:method}, in a dynamically
constrained inference, the eddy energy growth rate will be arrested by the
dynamical slumping of isopycnals, so that the resulting values of
$\lambda^{-1}(\phi,\theta)$ would not be as small, and reinforces the
interpretation that our inferred solution should be seen as a lower bound.

\begin{figure}
  \includegraphics[width=\textwidth]{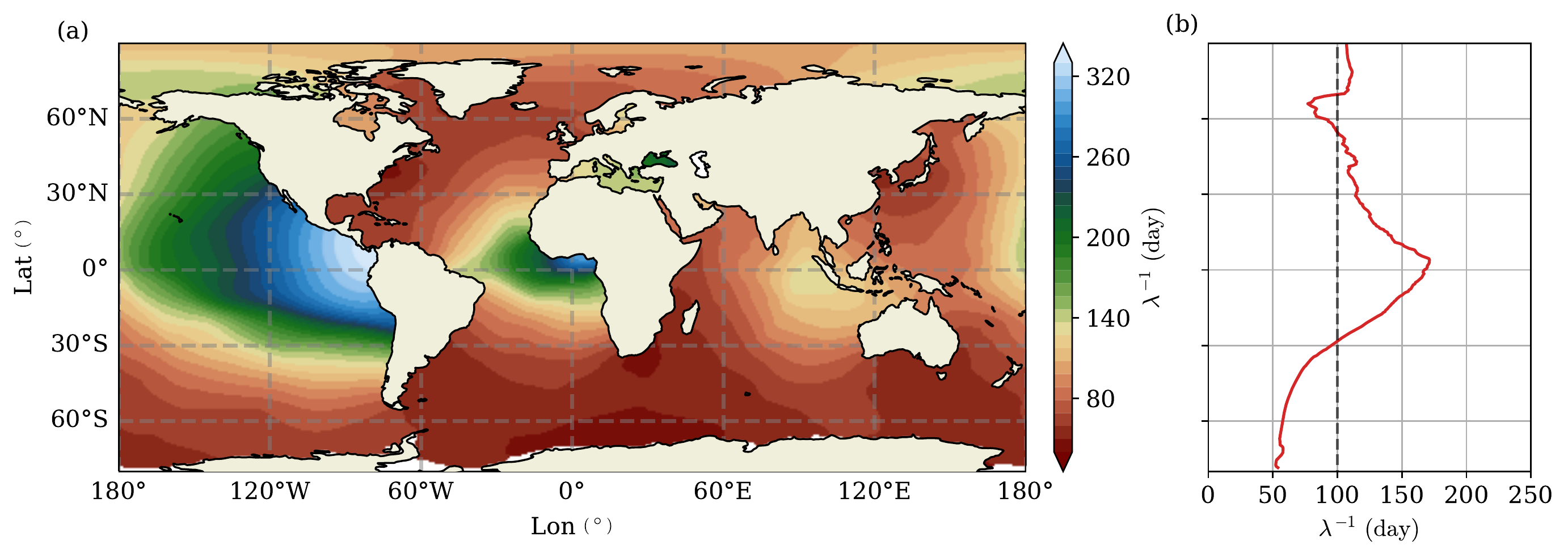}
  \caption{Optimized $\lambda^{-1}(\phi,\theta)$ for the $\epsilon = 5\times
  10^{15}$ calculation. ($a$) $\lambda^{-1}(\phi,\theta)$ in units of days.
  ($b$) the zonal average of $\lambda^{-1}(\phi,\theta)$, with the constant
  value of $\lambda^{-1} = 100$ days as used in the prognostic calculations of
  \citeA{Mak-et-al22} is marked as the dashed gray line.}
  \label{fig:lam_opt_eps5e15}
\end{figure}

The eddy energy dissipation time-scale $\lambda^{-1}(\phi, \theta)$ for the
control calculation is shown in Fig.~\ref{fig:lam_opt_eps5e15}. With regards to
spatial distribution, the values are small (i.e. short dissipation time-scale)
within the Southern Ocean (particularly in the Atlantic and Indian sectors),
around the Western Boundary Currents, and on the western ocean-land boundaries.
On the other hand, the values are large (i.e. long dissipation time-scale)
generally in the equatorial regions, but more wide spread in the Eastern
Pacific. The geographical locations of short eddy energy dissipation time-scale
are perhaps not surprising, given that these are regions with strong flows and
vigorous baroclinic eddy activity (cf. Fig.~\ref{fig:Eerr_eps5e15}$a$ in the
depth-integrated total eddy energy signature). The Southern Ocean and the
Western Boundary Currents are strongly turbulent regions, with significant mean
flows in the presence of rough bathymetry, leading to large eddy energy
dissipation via the multitude of dynamical processes given in \S\ref{sec:intro}.
The western boundary intensification of eddy energy dissipation is consistent
with the findings of \citeA{Zhai-et-al10}, resulting from eddy energy
convergence at the western boundaries via propagation of eddies at the long
Rossby wave phase speed \cite{Chelton-et-al11, KlockerMarshall14}, with energy
being transferred out of the geostrophic scales by processes such as loss of
balance and non-propagating form drag \cite{ZYang-et-al21}. The long dissipation
time-scales observed in the tropical regions and in the Eastern Pacific is also
consistent with expectations, as these are regions where baroclinic instability
is not expected to be particularly prevalent.

\begin{figure}
  \includegraphics[width=\textwidth]{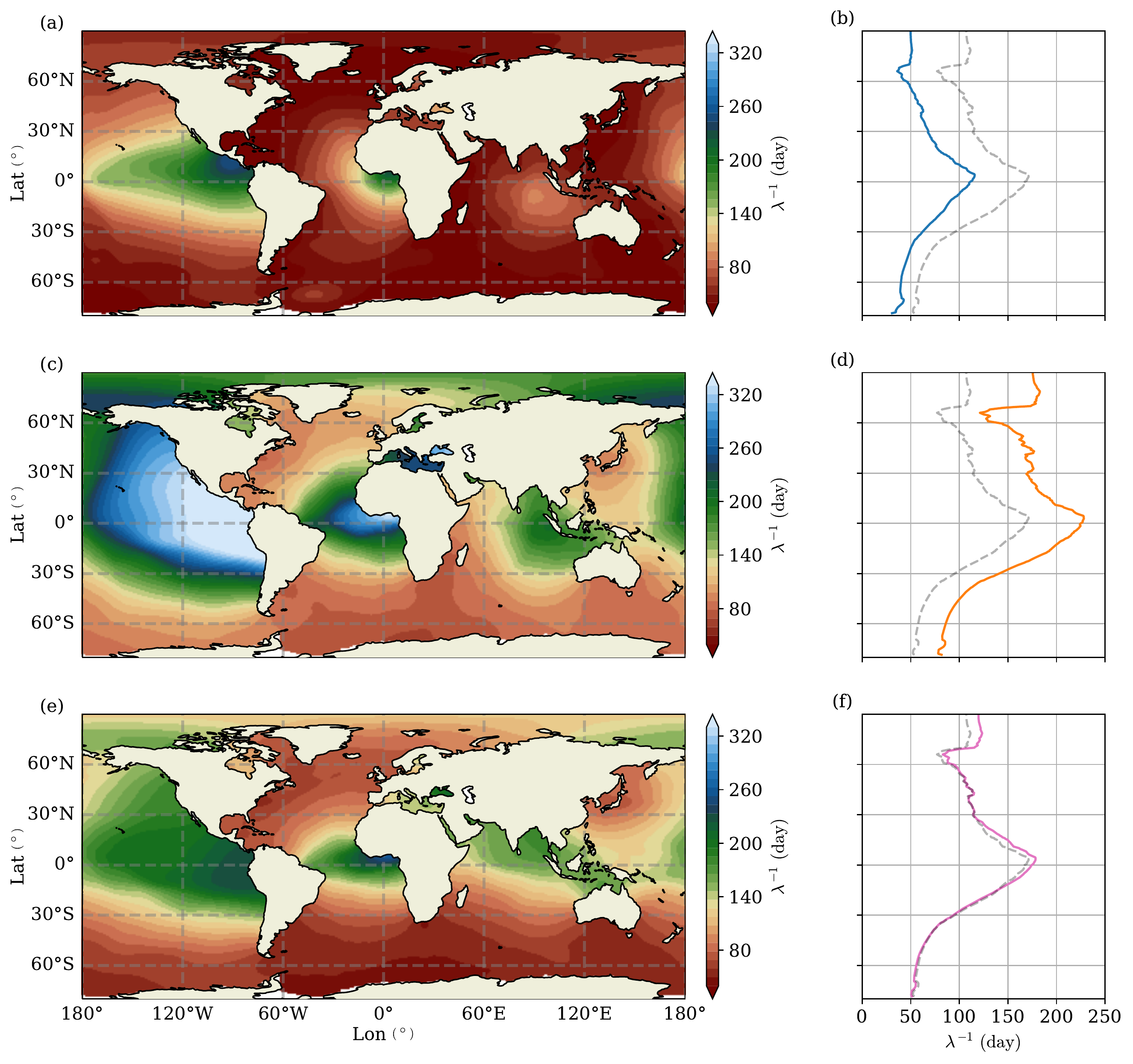}
  \caption{($a,c,e$) Inferred $\lambda^{-1}(\phi,\theta)$ (with red denoting
  larger eddy energy dissipation) and ($b,d,f$) its zonal average, where the
  dashed gray line denotes the control calculation $\epsilon=5\times10^{15}$.
  ($a,b$) $\epsilon=1\times10^{14}$; ($c,d$) $\epsilon=5\times10^{16}$; ($e,f$)
  $\epsilon=5\times10^{15}$ but with no advection contributions.}
  \label{fig:lam_vary_eps5e15}
\end{figure}

Fig.~\ref{fig:lam_vary_eps5e15}$a$--$d$ shows the representative features in the
inferred $\lambda^{-1}(\phi, \theta)$ for varying $\epsilon$. While calculations
at different values of $\epsilon$ possess qualitatively similar spatial
distributions, for lower values of $\epsilon$, the solution is allowed to take
more extreme values and the resulting values of $\lambda^{-1}(\phi,\theta)$ is
generally smaller, while the converse is true.
Fig.~\ref{fig:lam_vary_eps5e15}$e,f$ shows the result of a calculation where we
take $\epsilon=5\times10^{15}$ but exclude the advective contribution in the
inference calculation (by commenting out the last line defining \verb|F| in
Fig.~\ref{fig:code} when performing the optimization calculation). The resulting
spatial and zonal distribution is largely similar to the solutions with
advection, but with reduced western intensification, particularly in the Western
Pacific, Indian subtropics and Eastern Australia, attributed to lack of eddy
energy advection westward at the long Rossby phase speed. The inclusion of
advection leads to local differences in the distribution of the eddy energy and
thus $\lambda^{-1}(\phi, \theta)$, but with only minor difference to the overall
magnitudes in its zonal average.

The parameters chosen for the calculations given in Table \ref{tbn:param} were
motivated by the choices made in prognostic calculations of \cite{Mak-et-al22}
with NEMO ORCA1 employing GEOMETRIC, though we are free to explore the parameter
space for the inference calculation. It is found that if we increase the
horizontal eddy energy diffusion coefficient $\eta_E$ for fixed $\epsilon$, the
model given by Eq.~(\ref{eq:model-weak1}) outputs a $\hat{E}$ that is weaker and
more diffused, and the optimizer returns a $\lambda^{-1}(\phi,\theta)$ with
similar spatial distributions but marginally larger values. Increasing
$\hat{E}_0$ by no more than an order of magnitude for fixed $\epsilon$ leads to
very mild effects in the basin regions where $\hat{E}$ is small. Increasing
$\alpha$ for fixed $\epsilon$ leads to globally smaller values of
$\lambda^{-1}(\phi,\theta)$, since $\alpha$ affects the growth term in
Eq.~(\ref{eq:model-weak1}), and thus calls for larger eddy energy dissipation to
maintain balance. Beyond the eddy energy equation parameters, if $\epsilon
\lesssim 10^{13}$, the optimization calculation starts to have large variations
in $\lambda^{-1}(\phi,\theta)$ and can return negative values of $\hat{E}$; such
results were deemed unphysical and have been omitted from this work. While we
employ a finite element mesh with a horizontal characteristic length scale of
$100$ km, changing the mesh resolution itself has very little influence on the
solutions as long as $\epsilon$ is larger than about $10^{14}$, but with a
significant increase in computation cost mostly from memory requirements
(through the elliptic solve rather than the L-BFGS-B algorithm). The reason for
the relative invariance of the solution to the finite element mesh resolution is
from the fact that the current inference problem is diagnostic in nature, and it
is the resolution of the model generating the forcing data for the inverse
calculation that has the most effect.

%-------------------------------------------------------------------------------

\section{Prognostic model calculations}\label{sec:prognostic}

The presence of feedback loops means there is no guarantee that a diagnostic
result such as the one here will lead to improvements in a prognostic
calculation. The inferred $\lambda^{-1}(\phi, \theta)$ is thus utilized as a
prescribed input is utilized in an ocean global circulation model in prognostic
mode, to assess the consequences on the model output and behavior compared with
the case where a prescribed spatially constant $\lambda^{-1}$ field is used
\cite<cf.>{Mak-et-al22}. The principal hypothesis is that the use of
$\lambda^{-1}(\phi, \theta)$ will improve on the resulting parameterized total
eddy energy signature. A secondary aim is a demonstration that the inferred
$\lambda^{-1}(\phi, \theta)$ is a physically plausible lower bound for the
mesoscale eddy energy dissipation time-scale.

The model and set up we employ is exactly the same as one in
\citeA{Mak-et-al22}. An ocean only global configuration model employing NEMO
\cite[v3.7dev r8666]{Madec-NEMO} using a tri-polar grid ORCA1 grid
\cite{MadecImbard96} with nominally $1^\circ$ horizontal resolution (ORCA1) with
the LIM3 ice model \cite{Rousset-et-al15} is utilized. With 46 uneven vertical
levels, placing more resolution near the ocean surface, the model employs the
TEOS-10 equation of state \cite{Roquet-et-al15a}. Forcing by the atmosphere is
modeled by the NCAR bulk formulae with normal year forcing \cite{LargeYeager09}.
Sea surface salinity but not temperature restoration is included to reduce model
drift. The GEOMETRIC implementation in NEMO for this model is described in
\citeA{Mak-et-al22}, and is essentially given by Eq.~(\ref{eq:GEOMloc-e}) here,
with the addition of the $E_0$ term (cf. Eq.~\ref{eq:model-weak1}). The
parameters chosen for the prognostic calculations are exactly those given in
Table~\ref{tbn:param}. The model is spun up from WOA13 climatology
\cite{Locarini-et-al13, Zweng-et-al13} for 300 years. Prognostic calculations
with $\lambda^{-1}(\phi, \theta)$ for $\epsilon = 5\times 10^{15}$ and $\epsilon
= 5\times 10^{16}$ (cf. Fig.~\ref{fig:lam_opt_eps5e15} and
\ref{fig:lam_vary_eps5e15}$b$) are reported here; the calculation with
$\lambda^{-1}(\phi, \theta)$ for $\epsilon=1\times10^{14}$ (cf.
Fig.~\ref{fig:lam_vary_eps5e15}$a$) was found to be numerically unstable in the
parameterized eddy energy equation and has been omitted.

\begin{figure}
  \includegraphics[width=\textwidth]{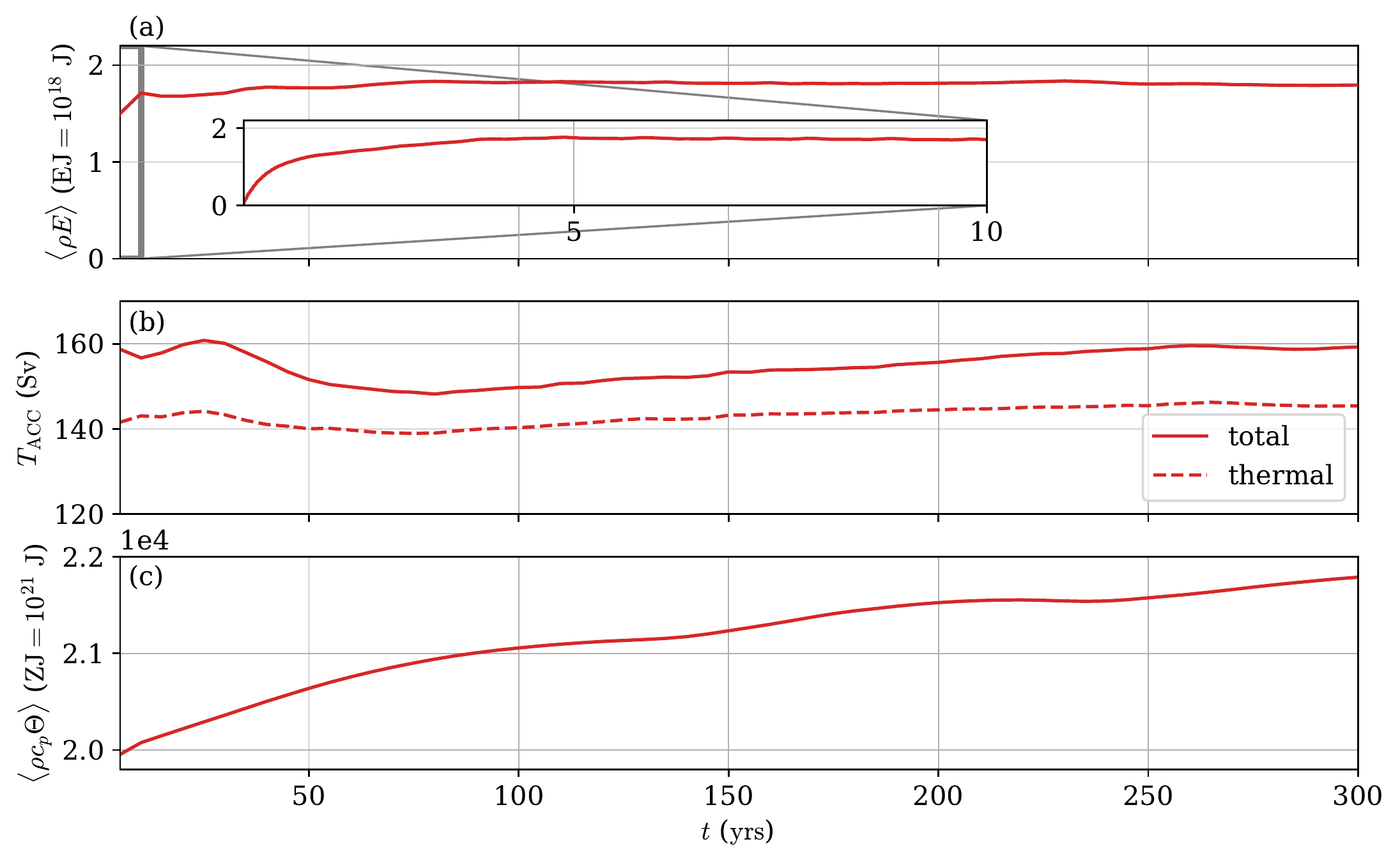}
  \caption{Time-series of the prognostic calculation employing the inferred
  $\lambda^{-1}(\phi, \theta)$ with $\epsilon = 5\times 10^{15}$. ($a$) Domain
  integrated eddy energy $\langle \rho E\rangle$ from GEOMETRIC (in units of
  $\mathrm{EJ} = 10^{18}\ \mathrm{J}$, with $\rho_0=1026\ \mathrm{kg}\
  \mathrm{m}^{-3}$), with inset showing the same quantity at higher temporal
  resolution over the first 10 years. ($b$) Antarctic Circumpolar Current
  transport $T_{\rm ACC}$ with total and thermal wind component (in units of
  Sverdrups), where thermal wind component defined as the residual of the total
  minus the depth-integrated bottom flow transport. ($c$) Domain-integrated
  ocean heat content $\langle\rho c_p \Theta\rangle$, where $\rho$ is the
  locally referenced density, $c_p$ is the heat capacity, and $\Theta$ is the
  conservative temperature.}
  \label{fig:ORCA1_time_series}
\end{figure}

Denoting $\langle\cdot\rangle$ to be the domain-integrated quantity, we first
show in Fig.~\ref{fig:ORCA1_time_series} the time-series of the calculation
utilizing $\lambda^{-1}(\phi, \theta)$ with $\epsilon = 5\times 10^{15}$ for the
diagnosed domain-integrated eddy energy $\langle \rho E\rangle$, the total and
thermal wind component of the Antarctic Circumpolar Current transport $T_{\rm
ACC}$ (where the total is the transport over the whole depth of the Drake
passage, and the thermal wind component is calculated as the residual of the
total and the analogous depth-integrated bottom flow), and the total ocean heat
content $\langle \rho c_p \Theta\rangle$, where $\rho$ is the locally referenced
density, $c_p$ is the heat capacity, and $\Theta$ is the conservative
temperature. The main purpose here is to demonstrate the multiple adjustment
time-scales inherent in the different diagnosed quantities. The parameterized
eddy energy as represented through GEOMETRIC here adjusts on a fast time-scale
(of around 5 to 10 years), while the $T_{\rm ACC}$ might be argued to have
reached a quasi-equilibrium over centennial time-scales but, like the ocean heat
content, there is a much longer adjustment on millennium time-scales associated
with the deep/abyssal stratification \cite<e.g.>{Mak-et-al22}.

With these observations in mind, for simplicity and computation resource
reasons, we will make a direct comparison of the present calculations employing
$\lambda^{-1}(\phi, \theta)$ where diagnostics were averaged over the model
years 290 to 300, with the previously reported results of \citeA{Mak-et-al22}
employing a spatially constant $\lambda^{-1} = 100$ days that already exists,
but where the data was diagnosed from averages over the model years 3000 to
3100. For comparing the eddy energy signature this should not be an issue, given
both calculations have a parameterized (domain-integrated and spatially varying)
eddy energy signature that will have equilibrated. For comparing circulation
metrics such as $T_{\rm ACC}$ and the Atlantic Meridional Overturning
Circulation $T_{\rm AMOC}$, the diagnosed values can be compared as long as we
bear in mind that there will likely be 10 to 20\% upward drift of $T_{\rm ACC}$
and $T_{\rm AMOC}$ in the present results until equilibration occurs over the
500 to 1000 year time-scale. We refrain from comparing the ocean heat content.

To quantify whether our choice of $\lambda^{-1}(\phi,\theta)$ is able to improve
the eddy energy signature, we take $\hat{E}_{\rm data}$ as diagnosed from
ORCA0083-N01 (interpolated onto the ORCA1 grid) as the reference, and compute
the area-averaged $L^2$ mismatch of the corresponding $\hat{E}_{\rm ORCA1}$ from
the prognostic calculations, given by
\begin{equation}\label{eq:J3} J_3 =
  \frac{1}{A}\|\hat{E}_{\rm data} - \hat{E}_{\rm ORCA1}\|^2_{L^2} = \frac{1}{A}\int_A \left(\hat{E}_{\rm data} - \hat{E}_{\rm ORCA1}\right)^2\; \mathrm{d}A. 
\end{equation}
We additionally compute the domain-integrated total eddy energy value $\langle
\rho E\rangle$, the total and thermal wind component of $T_{\rm ACC}$, and the
Atlantic Meridional Overturning Circulation $T_{\rm AMOC}$ (diagnosed as the
northward transport over the top 1000 m at $26^\circ\ \textnormal{N}$ at the
Western side of the Atlantic). The computed diagnostics are given in
Table~\ref{tbn:diags}.

\begin{table}
  \caption{Computed diagnostics of the calculations employing spatially constant
  $\lambda^{-1}$ \cite<100 days, data from>{Mak-et-al22} and calculations
  employing the inferred $\lambda^{-1}(\phi, \theta)$ for $\epsilon =
  5\times10^{15}$ and $5\times10^{16}$ (the prognostic calculation for $\epsilon
  = 1\times10^{14}$ was numerically unstable and omitted). Diagnostics are:
  $J_3$, the area averaged mismatch of the depth-integrated eddy energy as
  compared to the analogous ORCA0083-N01 reference calculation; $\langle\rho
  E\rangle$, the domain-integrated eddy energy from the prognostic calculations;
  $T_{\mathrm{ACC}}$, the Antarctic Circumpolar Current transport (total and
  thermal wind component), diagnosed as the transport over the model Drake
  passage; $T_{\mathrm{AMOC}}$ diagnosed as the northward transport over the top
  1000 m at $26^\circ\ \textnormal{N}$ at the Western side of the Atlantic. For
  reference, ORCA0083-N01 has $\langle\rho E_{\rm data}\rangle = 9.52\
  \mathrm{EJ}$.}
  \label{tbn:diags}
  \centering
  \begin{tabular}{l l l l l}
  \hline
   & $J_3$ ($\mathrm{m}^3\ \mathrm{s}^{-2}$) & $\langle\rho E\rangle$ ($\mathrm{EJ} = 10^{18}\ \mathrm{J}$) & $T_{\mathrm{ACC}}$ (Sv) & $T_{\mathrm{AMOC}}$ (Sv) \\
  \hline
   $\lambda^{-1} = 100\ \mathrm{days}$ & $1922.22$ & $3.33$ & $138.10$ (total) & $10.38$\\
   \cite{Mak-et-al22} & & & $121.23$ (thermal) \\
   \hline
   $\lambda^{-1}(\theta,\phi)$ & $1560.89$ & $1.80$ & $159.17$ (total) & $11.13$\\
   $(\epsilon=5\times 10^{15})$ & & & $145.41$ (thermal) \\
    \hline
   $\lambda^{-1}(\theta,\phi)$ & $1954.70$ & $2.98$ & $143.27$ (total) & $9.05$\\
   $(\epsilon=5\times 10^{16})$ & & & $99.29$ (thermal) \\
  \hline
\end{tabular}
\end{table}

\begin{figure}
  \includegraphics[width=\textwidth]{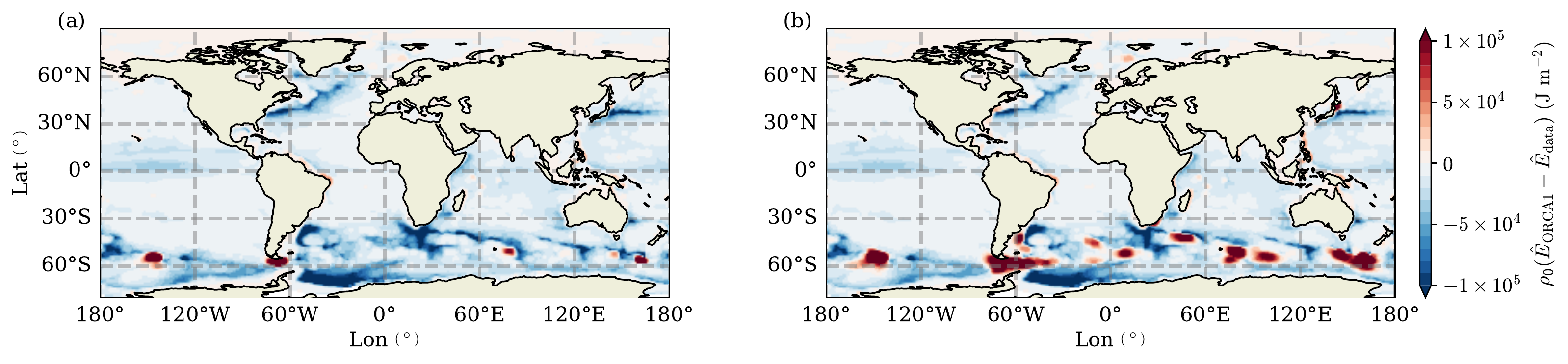}
  \caption{Spatial distribution of the signed mismatch $\rho_0(\hat{E}_{\rm
  ORCA1} - \hat{E}_{\rm data})$, where $\rho_0=1026\ \mathrm{kg}\
  \mathrm{m}^{-3}$, so data is now in units of $\mathrm{J}\ \mathrm{m}^{-2}$.
  Both simulations employ the standard normal year forcing with no amplified
  wind, with ($a$) utilizing $\lambda^{-1}(\phi,\theta)$ with
  $\epsilon=5\times10^{15}$ (cf. Fig.~\ref{fig:lam_opt_eps5e15}), and ($b$)
  $\lambda_0^{-1} = 100\ \textnormal{days}$ everywhere.}
  \label{fig:ORCA1_Eerr}
\end{figure}

From the computed values of $J_3$, it may be seen that the calculation using
$\lambda^{-1}(\phi, \theta)$ with $\epsilon = 5\times 10^{15}$ has a lower $L^2$
mismatch compared to the calculation with $\epsilon = 5\times 10^{16}$ and the
one with spatially constant $\lambda^{-1}$, although arguably the improvements
are somewhat modest. Fig.~\ref{fig:ORCA1_Eerr} shows the spatial distribution of
the signed mismatch, and the lower values of $J_3$ from the prognostic
calculations using $\lambda^{-1}(\phi, \theta)$ with $\epsilon = 5\times
10^{15}$ compared to the calculation with $\lambda_0^{-1} = 100\
\textnormal{days}$ results primarily come from improvements within the Southern
Ocean, where the prognostic calculations using $\lambda^{-1}(\phi,\theta)$ has
reduced coverage of positive biases in the depth-integrated eddy energy. The
reduction in the average values of the eddy energy $\langle \rho E\rangle$ can
also be seen to be arising from the reduced values of the eddy energy in the
Southern Ocean.

The negative biases however are generally large, as can be seen in
Fig.~\ref{fig:ORCA1_Eerr}. Additionally, note that the domain-integrated eddy
energy value of the reference ORCA0083-N01 calculation is $\langle\rho E_{\rm
data}\rangle = 9.52\ \mathrm{EJ}$, and the corresponding values from the
prognostic calculations given in Table~\ref{tbn:diags} are a few factors lower.
While the Southern Ocean biases have been reduced somewhat, over the rest of the
globe the negative bias is prevalent, which is somewhat consistent with an
eddy energy dissipation time-scale that is too short.

Beyond the response in the eddy energy signature, the calculations with
$\lambda^{-1}(\phi, \theta)$ also results in a plausible $T_{\rm ACC}$ as well
as $T_{\rm AMOC}$, although the latter is a little on the low side; see for
example Table 2 in \citeA{Farneti-et-al15} and Fig. 1 of
\citeA{Danabasoglu-et-al14} for a summary of model and observational estimates
for $T_{\rm ACC}$ and $T_{\rm AMOC}$ respectively. It should be noted that the
calculation using $\lambda^{-1}(\phi, \theta)$ with $\epsilon = 5\times 10^{15}$
has the highest $T_{\rm ACC}$ (both total and thermal wind component) and
$T_{\rm AMOC}$. This is consistent with expectations, since the associated
dissipation time-scale in the Southern Ocean is the shortest, leading to reduced
flattening of isopycnals by mesoscale eddies, steeper stratification in the
Southern Ocean \cite{Marshall-et-al17, Mak-et-al18}, and in turn a deepening of
the global pycnocline and increased $T_{\rm AMOC}$ via isopycnal connectivity
\cite{MarshallJohnson17}, something that has been demonstrated in numerical
models \cite{Mak-et-al18, Mak-et-al22}. The results here also provide some
additional evidence for the interpretation that the $\lambda^{-1}(\phi,\theta)$
given here should be viewed as a lower bound, since $T_{\rm ACC}$ is already on
the rather high side.

\begin{figure}
  \includegraphics[width=\textwidth]{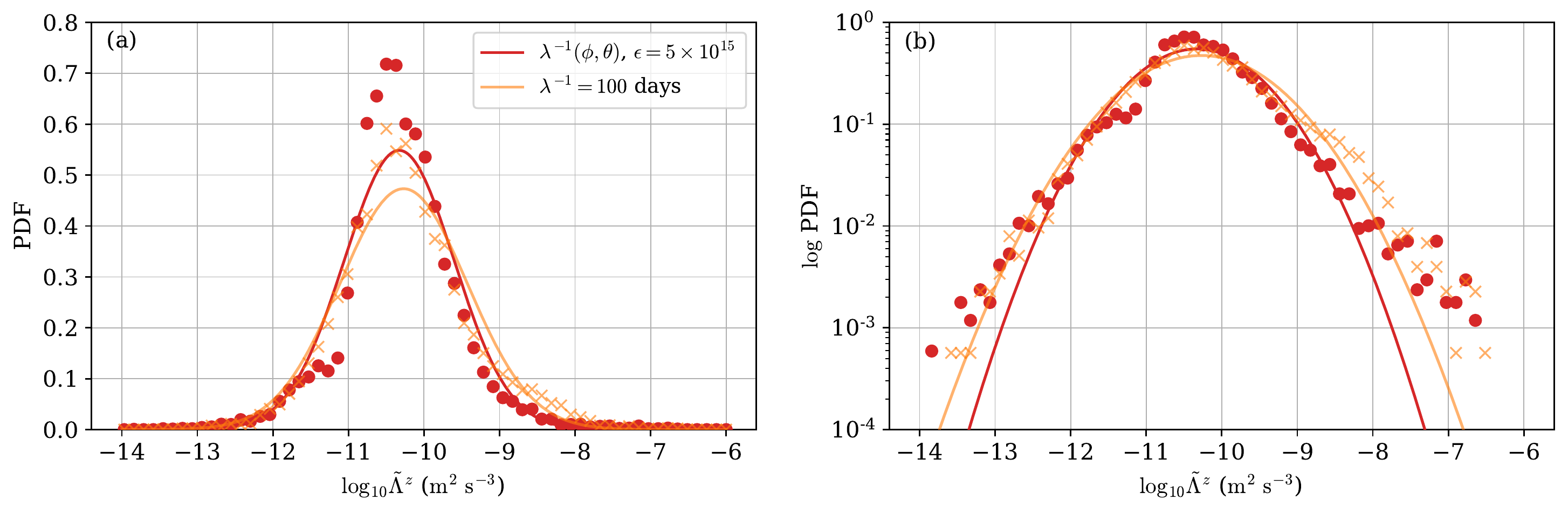}
  \caption{The PDF of the depth-averaged total eddy energy dissipation from the
  calculation utilizing $\lambda^{-1}(\phi,\theta)$ with
  $\epsilon=5\times10^{15}$ (red) and $\lambda_0^{-1} = 100\ \textnormal{days}$
  everywhere (orange), for diagnosed data (markers) and the best-fit Gaussian
  PDF (lines). The PDF shows one particular snapshot from five day averaged data
  within the averaging period of the respective prognostic calculation, but is
  representative of other sampled time snapshots as well as similar to the
  analogous PDF from time-mean data. ($a$) The PDF of the
  $\log_{10}\tilde{\Lambda}^z$ given in Eq.~(\ref{eq:diss_zavg}). ($b$) Same as
  ($a$) but on a base 10 logarithmic scale.}
  \label{fig:diss_pdf}
\end{figure}

It may be of interest to consider the statistical distribution in addition to
the spatial distribution of the eddy energy dissipation. We note that the work
of \citeA{PearsonFoxKemper18} reports that the dissipation of eddy kinetic
energy in a high resolution global ocean model follows a log-normal distribution
on horizontal slices, with slightly varying statistics in depth, and makes the
point that mesoscale parameterizations should be such that the energy flux
statistics are consistent with the reported diagnostics. We show in
Fig.~\ref{fig:diss_pdf} the probability distribution function (PDF) of the
depth-averaged eddy energy dissipation as parameterized, given by
\begin{equation}\label{eq:diss_zavg}
  \tilde{\Lambda}^z = \frac{1}{H} \left(\lambda \int^H_0 (E - E_0)\; \mathrm{d}z\right),
\end{equation}
which has units of $\mathrm{m}^{2}\ \mathrm{s}^{-3}$ like the turbulent energy
flux in \citeA{PearsonFoxKemper18}, for the prognostic calculation using
$\lambda^{-1}(\theta, \phi)$ with $\epsilon = 5\times 10^{15}$ as well as for
the case with uniform $\lambda^{-1}=100$ days.

A direct comparison of the results here with the work of
\citeA{PearsonFoxKemper18} is slightly problematic in the sense that the
quantities to be compared are different (depth-integrated dissipation of eddy
kinetic and potential energy here as compared to horizontal slices of eddy
kinetic energy dissipation), but nevertheless the values of the energy
dissipation obtained here are not unreasonable compared to Fig.~1 of
\citeA{PearsonFoxKemper18}, although our dissipation is on the larger side.
While there are hints of log-Gaussianity in our PDFs, there is also notable
skewness to the larger values, as well and kurtosis in the data reflected by the
heavy tails, as can be seen in Fig.~\ref{fig:diss_pdf}$b$ with the distribution
of data points relative to the best fit log-Gaussian PDF. The use of
$\lambda^{-1}(\phi,\theta)$ with $\epsilon=5\times10^{15}$ additionally leads to
a reduction in the variance and larger dissipation values compared to the case
with spatially uniform $\lambda^{-1}$, as seen by the relatively more narrow
best-fit PDF, and a noticeable dip in the values of the data points at the
higher end of the distribution. The observed relative differences may be
attributed to a significant decrease in the positive biases of eddy energy
signatures in the Southern Ocean where dissipation is strong. Our interpretation
that $\lambda^{-1}(\phi, \theta)$ is expected to be a lower bound in the
time-scale is somewhat consistent with the observation that our dissipation rate
is possibly on the larger side compared to the values reported in
\citeA{PearsonFoxKemper18}.

%-------------------------------------------------------------------------------

\section{Conclusions and outlooks}\label{sec:conclusion}

In this work we provide a leading order constraint for the spatial distribution
of eddy energy flux out of the geostrophic mesoscales, interpreted here as a
geostrophic mesoscale eddy energy dissipation time-scale
$\lambda^{-1}(\phi,\theta)$. The problem is viewed as one of parameter inference
for $\lambda^{-1}(\phi,\theta)$, here inferred from high resolution numerical
model data and constrained by a parameterized eddy energy equation, as a
precursor for the more complete dynamically constrained inference problem
utilizing ocean observational data, to be discussed below. A simple and
computationally inexpensive optimization problem was performed, seeking an
optimal $\lambda^{-1}(\phi,\theta)$ that minimizes the mismatch between the
depth-integrated total eddy energy from the parameterized eddy energy equation
$\hat{E}(\lambda^{-1})$ and the depth-integrated total eddy energy diagnosed
from a high resolution numerical model $\hat{E}_{\rm data}$. The present
implementation utilizes the Firedrake software, leveraging the inbuilt solvers
as well as the automatic code generation capabilities to solve and explore the
associated optimization problem and its sensitivities to parameter choices. The
inferred $\lambda^{-1}(\phi,\theta)$ has the smallest values within the Southern
Ocean, Western Boundary Currents, and is western boundary intensified, regions
where baroclinic turbulence is particularly dominant, and coinciding with where
we expect the greatest energy flux out of the geostrophic mesoscales from
dynamical considerations \cite<e.g.>{Zhai-et-al10, NikurashinFerrari11,
Melet-et-al15, Rocha-et-al18, Rai-et-al21}. We caveat that while the inferred
spatial distribution of $\lambda^{-1}(\phi,\theta)$ may be consistent with
expectations, the resulting magnitudes should be viewed as a physically
plausible lower bound.

Prognostic calculations utilizing the inferred $\lambda^{-1}(\phi,\theta)$ in
the coarse resolution global configuration ocean model NEMO ORCA1 were
performed. The coarse resolution calculations result in an improved mismatch in
the parameterized depth-integrated eddy energy signature in the globally
integrated $L^2$ sense compared to a previous work that employs a spatially
constant $\lambda^{-1} = 100\ \textnormal{days}$ also in NEMO ORCA1
\cite{Mak-et-al22}, where the reference eddy energy signature is diagnosed from
the high resolution calculation NEMO ORCA0083-N01. The use of
$\lambda^{-1}(\phi,\theta)$ reduces positive biases in total eddy energy
signature in the Southern Ocean, though negative biases remain prevalent
throughout the globe, and the coarse resolution calculation possesses an average
eddy energy value that is too low. The diagnosed probability distribution for
the parameterized total eddy energy dissipation also seems to have a low bias,
compared to the diagnosed dissipation of eddy kinetic energy in
\citeA{PearsonFoxKemper18} from a high resolution ocean global circulation
model. The resulting circulation metrics from utilizing
$\lambda^{-1}(\phi,\theta)$ inferred here, such as the total and thermal wind
component of the Antarctic Circumpolar Current, is on the high side \cite<at
around 160 Sv, cf. Table 2 of>{Farneti-et-al15}, which is in line with the
dynamical arguments provided by \citeA{Marshall-et-al17} that a higher eddy
energy dissipation rate (so a shorter eddy energy dissipation time-scale) leads
to increased Antarctic Circumpolar Current transport and steepening of
isopycnals. Together, there is evidence in support of the inferred
$\lambda^{-1}(\phi,\theta)$ leading to an improved eddy energy signature in
prognostic calculations, as well as being a physically plausible lower bound for
the eddy energy dissipation time-scale.

Beyond providing a leading order constraint and estimate for eddy energy
dissipation time-scale $\lambda^{-1}(\phi,\theta)$ as progress towards
understanding ocean energetic pathways, this work also highlights and
demonstrates some perhaps lesser known but really quite powerful machinery
relating to inverse methods and calculations, such as automatic code generation
software such as Firedrake \cite{Rathgeber-et-al17}, automatic adjoint
generation libraries \cite<e.g.>{Farrell-et-al13, Maddison-et-al19} and mesh
generating software \cite{Avdis-et-al18} that is expected to have applications
in various branches of earth system modeling. Such tools have been applied to
problems in global tidal modeling with uncertain bathymetry
\cite{David-et-al19}, uncertainty quantification associated with ice sheets
\citeA{Kolziol-et-al21}, Tsunami source inversion \cite{Wallwork-thesis}, and
sediment transport modeling \cite{Clare-et-al22}, to name a few examples.

In order to obtain the present lower bound for $\lambda^{-1}(\phi,\theta)$, some
approximations were made in order to make the problem tractable. The
approximations we made were (1) prescribing the stratification and thus removing
dynamical feedbacks, and (2) the choice and assumption of data, cost functional,
control variable and/or constraining model. As argued in the methodology
section, the choice of prescribing the stratification, while reducing the
complexity of the inference problem (e.g. removing coupling to an ocean global
circulation model), has the consequence that the dynamical feedbacks are
removed. As a consequence, the reduction in the complexity, we are somewhat
restricted to the choice of target data, in this case to the total eddy energy
signature as the state variable. Given the reduced amount of target data, it
means we are somewhat limited to the choice and number of control variables we
can take unless we impose severe and somewhat ad hoc regularizations.

Ultimately, those were choices we made in light of our primary objective, to
provide a leading order reference constraint for which further investigations
can be based on. A ``simple'' fix in principle is to dispense with prescribing
the stratification parameters, and carry out a dynamically constrained inference
calculation, which was part of the motivation behind the present work. The
related machinery is already in place in the form of the ECCO framework within
MITgcm \cite<e.g.>{Forget-et-al15a, Fukumori-et-al18}, set up utilizing the
inbuilt algorithmic differentiation capabilities \cite<e.g>{GieringKaminski98}
to deriving adjoints for performing state estimation using the variational or
smoothing approach \cite<e.g.>{Kalnay-DA}. There is already a form of GEOMETRIC
in MITgcm \cite<from>{Mak-et-al18}, and the principal modification that would be
required for the intended dynamical inference is to couple the GEOMETRIC
parameterization for $\kappa_{\rm gm}$ accordingly to the existing ECCO
framework, and including $\lambda^{-1}(\phi, \theta)$ and possibly $\alpha$ as
additional control variables. By increasing the complexity of the problem, we
are now also able to utilize ocean observational data as the target instead of
just relying on the eddy energy signature. Given the large amounts of degrees of
freedom associated with the proposed control variables, the adjoint method,
which has a linear scaling in the complexity and requires a smaller amount of
model calculations \cite<one forward and one backward for each
iteration;>{Kalnay-DA, Gunzburger-control}, is a particularly well-suited
computational method \cite<cf. Green's function methods, while linear in
complexity, requires ocean general circulation model runs scaling with the
number of degrees of freedom, and is better when the number of control variables
are low; e.g.>{Nguyen-et-al11}. While ``simple'' in principle, in practice the
above proposal is still a formidable technical challenge and computationally
expensive, and given there have been no strong constraints on
$\lambda^{-1}(\phi, \theta)$ thus far, the present result serves as an important
leading order prior for the proposed dynamical inference calculation. The
proposed work is currently underway and will be reported in a future
publication.

A simpler procedure we have also considered is to stick with the strategy in
this work, but diagnose the \emph{time-varying} stratification and eddy energy
data from a high resolution model, and consider an adjoint-based calculation
where the cost functional is taken to be the mismatch between the parameterized
eddy energy and target eddy energy over time. The inference will still not be
dynamically constrained in the sense that the dynamical parameters for the
proposed calculation will not be functions of the state variable, i.e. changes
in $\lambda^{-1}(\phi,\theta)$ and thus $\hat{E}$ will have no bearing on the
evolution of the stratification, given the latter is prescribed. While the
calculation is certainly possible with the existing machinery since there are
adjoint libraries that can be coupled to Firedrake \cite<e.g.>{Farrell-et-al13,
Maddison-et-al19}, we are of the opinion that there is very little to be gained
from that approach, given the theoretical and probably practical limitation is
still that of prescribed stratification.

To close, we note that the variational methods considered here can be
interpreted in the Bayesian as a maximum likelihood approach
\cite<e.g.>{Kalnay-DA, BuiThanh-et-al13}, which by itself does not provide
estimates of the uncertainties, and can have issues with over-tuning in cases
where multiple different parameters are tuned at the same time
\cite{Williamson-et-al17}. Being able to quantify uncertainties associated with
inference calculations will also be a research focus in the planned
investigations. As an aside, sample prognostic calculations varying the Southern
Ocean wind stress \cite<cf.>{Mak-et-al22} employing the spatially varying eddy
energy dissipation time-scale appear to reproduce the eddy saturation phenomenon
in the Antarctic Circumpolar Current transport in the thermal wind component
\cite<e.g.,>{Munday-et-al13, Farneti-et-al15}, and shows hints of eddy
compensation \cite<e.g.>{GentDanabasoglu11, ViebahnEden12, Bishop-et-al16},
although limitation in available computational resources in this case did not
allow us to explore the sensitivity aspect fully. It is also of note that the
change in the magnitude of the eddy energy dissipation time-scale in prognostic
calculations, though relatively modest from a raw magnitude point of view, has a
rather large effect on the resulting ocean state (see Table \ref{tbn:diags}). In
that sense, the results here reinforce the conclusions of \citeA{Mak-et-al22}
that the ocean circulation displays a significant sensitivity to the eddy energy
dissipation time-scale, even on the centennial time-scales. A thorough
investigation into the sensitivity and benefits afforded by a more complex
representation of the eddy energetics in prognostic calculations \cite<e.g.,
with the Ocean Model Intercomparison Project (OMIP) protocol;>{Griffies-et-al16}
is beyond the scope of the present work, but is ongoing and will be reported in
due course.

%%%%%%%%%%%%%%

\section*{Data Availability}

For the pre-processing, NEMO ORCA0083-N01 data was obtained from
\url{http://gws-access.jasmin.ac.uk/public/nemo/}, provided by UK National
Oceanography Center through the JASMIN service. The base version of CDFTOOLS was
taken from \url{https://github.com/meom-group/CDFTOOLS}. For the optimization
calculations, this work uses a natively installed version of Firedrake, with a
mesh generated via the \verb|Qmesh| package (\url{https://www.qmesh.org/}) via a
Docker image, with the Firedrake wrapper of \texttt{tlm\_adjoint}
(\url{https://github.com/jrmaddison/tlm_adjoint}). The post-processing analysis
uses standard Python packages. Modifications of CDFTOOLS, the relevant Python
scripts, the processed data from this work (in both the native HDF5 finite
element format as well as the various gridded versions), and documentation of
software, versions and its dependencies are available on
\url{http://dx.doi.org/10.5281/zenodo.6559892}.

\acknowledgments

This research was funded by both the RGC Early Career Scheme 2630020 and the
Center for Ocean Research in Hong Kong and Macau, a joint research center
between the Qingdao National Laboratory for Marine Science and Technology and
Hong Kong University of Science and Technology. JM would additionally like to
thank Xiaoming Zhai, James Maddison and David Marshall for various technical
comments on the various iterations of the present work.

%% ------------------------------------------------------------------------ %%
%% References and Citations

%%%%%%%%%%%%%%%%%%%%%%%%%%%%%%%%%%%%%%%%%%%%%%%
%
% \bibliography{<name of your .bib file>} don't specify the file extension
%
% don't specify bibliographystyle
%%%%%%%%%%%%%%%%%%%%%%%%%%%%%%%%%%%%%%%%%%%%%%%

\bibliography{refs}

%Reference citation instructions and examples:
%
% Please use ONLY \cite and \citeA for reference citations.
% \cite for parenthetical references
% ...as shown in recent studies (Simpson et al., 2019)
% \citeA for in-text citations
% ...Simpson et al. (2019) have shown...
%
%
%...as shown by \citeA{jskilby}.
%...as shown by \citeA{lewin76}, \citeA{carson86}, \citeA{bartoldy02}, and \citeA{rinaldi03}.
%...has been shown \cite{jskilbye}.
%...has been shown \cite{lewin76,carson86,bartoldy02,rinaldi03}.
%... \cite <i.e.>[]{lewin76,carson86,bartoldy02,rinaldi03}.
%...has been shown by \cite <e.g.,>[and others]{lewin76}.
%
% apacite uses < > for prenotes and [ ] for postnotes
% DO NOT use other cite commands (e.g., \citet, \citep, \citeyear, \nocite, \citealp, etc.).
%

\end{document}